\documentclass[twocolumn]{aastex7} 
\usepackage{physics}
\newcommand{\nfrbcat}{4539}
\newcommand{\nsources}{3641}
\newcommand{\nsilver}{14}
\usepackage[dvipsnames]{xcolor}
\newcommand{\nrepeaters}{30}
\newcommand{\ntotalrepeaters}{80}
\newcommand{\nclusters}{}
\newcommand{\catii}{Cat2}
\newcommand{\cati}{Cat1}
\newcommand{\dmcosmic}{\text{DM}_{\text{cosmic}}}
\newcommand{\dmunits}{\text{pc cm}^{-3}}

\newcommand{\pcc}{$P_{\text{CC}}$}

\usepackage{amsmath}
\usepackage{amssymb}
\usepackage{physics}

\defcitealias{2020Natur.587...54C}{CHIME/FRB Collaboration et~al.}
\defcitealias{R3periodicity}{CHIME/FRB Collaboration et~al.}
\defcitealias{RN1}{CHIME/FRB Collaboration et~al.}
\defcitealias{R2}{CHIME/FRB Collaboration et~al.}
\defcitealias{RN3}{CHIME/FRB Collaboration et~al.}
\defcitealias{rbfloat}{CHIME/FRB Collaboration et~al.}
\defcitealias{cat2}{CHIME/FRB Collaboration et~al.}

\begin{document}

\title{Discovery of 30 Repeating Fast Radio Burst Sources and Uniform Population Statistics of 80 Repeating Sources from CHIME/FRB}



\correspondingauthor{Amanda M. Cook}
\email{amanda.cook@mail.mcgill.ca}

\author[0000-0001-6422-8125]{Amanda M. Cook}
\email{amanda.cook@mail.mcgill.ca}
\affiliation{Department of Physics, McGill University, 3600 rue University, Montr\'eal, QC H3A 2T8, Canada}
\affiliation{Trottier Space Institute, McGill University, 3550 rue University, Montr\'eal, QC H3A 2A7, Canada}
\affiliation{Anton Pannekoek Institute for Astronomy, University of Amsterdam, Science Park 904, 1098 XH Amsterdam, The Netherlands}

\author[0000-0002-6823-2073]{Kaitlyn Shin}
\email{kaitshin@caltech.edu}
\affiliation{Cahill Center for Astronomy and Astrophysics, MC 249-17 California Institute of Technology, Pasadena CA 91125, USA}

\author[0000-0002-4795-697X]{Ziggy Pleunis}
\email{z.pleunis@uva.nl}
\affiliation{Anton Pannekoek Institute for Astronomy, University of Amsterdam, Science Park 904, 1098 XH Amsterdam, The Netherlands}
\affiliation{ASTRON, Netherlands Institute for Radio Astronomy, Oude Hoogeveensedijk 4, 7991 PD Dwingeloo, The Netherlands}

\author[0009-0006-1258-4228]{Maxwell Fine}
\email{maxwell.fine@mail.mcgill.ca}
\affiliation{Department of Physics, McGill University, 3600 rue University, Montr\'eal, QC H3A 2T8, Canada}

\author[0009-0009-0938-1595]{Naman Jain}
\email{naman.jain@mail.mcgill.ca}
\affiliation{Department of Physics, McGill University, 3600 rue University, Montr\'eal, QC H3A 2T8, Canada}
\affiliation{Trottier Space Institute, McGill University, 3550 rue University, Montr\'eal, QC H3A 2A7, Canada} 

\author[0000-0001-5628-7256]{Derek Bingham}
\email{dbingham@sfu.ca}
\affiliation{Department of Statistics and Actuarial Science, 8888 University Dr W, Burnaby, BC V5A 1S6, Canada}

\author[0000-0002-8376-1563]{Alice P. Curtin}
\email{alice.curtin@mail.mcgill.ca}
\affiliation{Department of Physics, McGill University, 3600 rue University, Montr\'eal, QC H3A 2T8, Canada}
\affiliation{Trottier Space Institute, McGill University, 3550 rue University, Montr\'eal, QC H3A 2A7, Canada}
\affiliation{Anton Pannekoek Institute for Astronomy, University of Amsterdam, Science Park 904, 1098 XH Amsterdam, The Netherlands}

\author[0000-0003-3734-8177]{Gwendolyn Eadie}
\email{gwen.eadie@utoronto.ca}
  \affiliation{David A.\ Dunlap Department of Astronomy and Astrophysics, 50 St. George Street, University of Toronto, ON M5S 3H4, Canada}
    \affiliation{Department of Statistical Sciences, University of Toronto, Toronto, ON M5S 3G3, Canada}
    \affiliation{Data Sciences Institute, University of Toronto, 17th Floor, Ontario Power Building, 700 University Ave, Toronto, ON M5G 1Z5, Canada}
\author[0000-0002-3382-9558]{B. M. Gaensler}
\email{gaensler@ucsc.edu}
\affiliation{Department of Astronomy and Astrophysics, University of California, Santa Cruz, 1156 High Street, Santa Cruz, CA 95060, USA}
\affiliation{Dunlap Institute for Astronomy and Astrophysics, 50 St. George Street, University of Toronto, ON M5S 3H4, Canada}
\affiliation{David A.\ Dunlap Department of Astronomy and Astrophysics, 50 St. George Street, University of Toronto, ON M5S 3H4, Canada}

\author[0000-0003-2317-1446]{Jason W.T. Hessels}
\email{jason.hessels@mcgill.ca}
\affiliation{Department of Physics, McGill University, 3600 rue University, Montr\'eal, QC H3A 2T8, Canada}
\affiliation{Trottier Space Institute, McGill University, 3550 rue University, Montr\'eal, QC H3A 2A7, Canada}
\affiliation{Anton Pannekoek Institute for Astronomy, University of Amsterdam, Science Park 904, 1098 XH Amsterdam, The Netherlands}
\affiliation{ASTRON, Netherlands Institute for Radio Astronomy, Oude Hoogeveensedijk 4, 7991 PD Dwingeloo, The Netherlands}

\author[0000-0002-4209-7408]{Calvin Leung}
\email{calvin_leung@berkeley.edu}
\affiliation{Miller Institute for Basic Research, Stanley Hall, Room 206B, Berkeley, CA 94720, USA}
 \affiliation{Department of Astronomy, University of California, Berkeley, CA 94720-3411, USA}

\author[0000-0002-7164-9507]{Robert Main}
\email{robert.main@mcgill.ca}
\affiliation{Department of Physics, McGill University, 3600 rue University, Montr\'eal, QC H3A 2T8, Canada}
\affiliation{Trottier Space Institute, McGill University, 3550 rue University, Montr\'eal, QC H3A 2A7, Canada}

\author[0000-0001-7491-046X]{Nicole Mulyk}
\email{nicole.mulyk@mail.mcgill.ca}
\affiliation{Department of Physics, McGill University, 3600 rue University, Montr\'eal, QC H3A 2T8, Canada}
\affiliation{Trottier Space Institute, McGill University, 3550 rue University, Montr\'eal, QC H3A 2A7, Canada}

\author[0000-0002-8897-1973]{Ayush Pandhi}
\email{ayush.pandhi@mcgill.ca}
\affiliation{Department of Physics, McGill University, 3600 rue University, Montr\'eal, QC H3A 2T8, Canada}
\affiliation{Trottier Space Institute, McGill University, 3550 rue University, Montr\'eal, QC H3A 2A7, Canada}

\author[0000-0002-7374-7119]{Paul Scholz}
\email{pscholz@yorku.ca}
\affiliation{Department of Physics and Astronomy, York University, 4700 Keele Street, Toronto, ON MJ3 1P3, Canada}

\author[0000-0003-2631-6217]{Seth R. Siegel}
\email{Seth.Siegel@skao.int}
\affiliation{SKAO, Science Operations Centre, CSIRO ARRC, 26 Dick Perry Avenue, Kensington WA 6151 Australia}
\affiliation{Perimeter Institute of Theoretical Physics, 31 Caroline Street North, Waterloo, ON N2L 2Y5, Canada}
\affiliation{Department of Physics, McGill University, 3600 rue University, Montr\'eal, QC H3A 2T8, Canada}
\affiliation{Trottier Space Institute, McGill University, 3550 rue University, Montr\'eal, QC H3A 2A7, Canada}

\author[0000-0002-9761-4353]{David C. Stenning}
\email{dstennin@sfu.ca}
\affiliation{Department of Statistics and Actuarial Science, 8888 University Dr W, Burnaby, BC V5A 1S6, Canada}

\author[0000-0001-5002-0868]{Thomas C. Abbott}
\email{thomas.abbott@mail.mcgill.ca}
  \affiliation{Department of Physics, McGill University, 3600 rue University, Montr\'eal, QC H3A 2T8, Canada}
  \affiliation{Trottier Space Institute, McGill University, 3550 rue University, Montr\'eal, QC H3A 2A7, Canada}
\author[0000-0001-5908-3152]{Bridget C Andersen}
\email{banders@ucsc.edu}
\affiliation{Department of Astronomy and Astrophysics, University of California, Santa Cruz, 1156 High Street, Santa Cruz, CA 95060, USA}

\author[0000-0002-3615-3514]{Mohit Bhardwaj}
\email{mohitb@iitk.ac.in}
\affiliation{Department of Space Planetary Astronomical Sciences and Engineering, Indian Institute of Technology KanpuR, Uttar Pradesh, India}
\author[0009-0001-0983-623X]{Alice Cai}
\email{}
\affiliation{Department of Physics and Astronomy, Northwestern University, Evanston, IL 60208, USA}
\affiliation{Center for Interdisciplinary Exploration and Research in Astronomy, Northwestern University, 1800 Sherman Avenue, Evanston, IL 60201, USA}

\author[0000-0002-2878-1502]{Shami Chatterjee}
\email{shami@astro.cornell.edu}
\affiliation{Cornell Center for Astrophysics and Planetary Science, Cornell University, Ithaca, NY 14853, USA}

\author[0000-0003-4098-5222]{Fengqiu Adam Dong}
\email{fadong@yorku.ca}
\affiliation{Department of Physics and Astronomy, York University, 4700 Keele Street, Toronto, ON MJ3 1P3, Canada}

\author[0000-0001-8384-5049]{Emmanuel Fonseca}
\email{emmanuel.fonseca@mail.wvu.edu}
\affiliation{Department of Physics and Astronomy, West Virginia University, PO Box 6315, Morgantown, WV 26506, USA}
\affiliation{Center for Gravitational Waves and Cosmology, West Virginia University, Chestnut Ridge Research Building, Morgantown, WV 26505, USA}

\author[0000-0002-5794-2360]{Danté M. Hewitt}
\email{d.m.hewitt@uva.nl}
\affiliation{Anton Pannekoek Institute for Astronomy, University of Amsterdam, Science Park 904, 1098 XH Amsterdam, The Netherlands}

\author[0000-0003-3457-4670]{Ronniy C. Joseph}
\email{ronniy.joseph@stcorp.nl}
\affiliation{S\&T Netherlands, Olof Palmestraat 14, 2616 LR Delft, Netherlands}
\author[0009-0007-5296-4046	]{Lordrick Kahinga}
\email{lkahinga@ucsc.edu}
\affiliation{Department of Astronomy and Astrophysics, University of California, Santa Cruz, 1156 High Street, Santa Cruz, CA 95060, USA}
\affiliation{Department of Physics, College of Natural and Mathematical Sciences, University of Dodoma, 1 Benjamin Mkapa Road, 41218 Iyumbu, Dodoma, Tanzania}
\author[0000-0002-5857-4264]{Mattias Lazda}
\email{mattias.lazda@mail.utoronto.ca}
\affiliation{David A.\ Dunlap Department of Astronomy and Astrophysics, 50 St. George Street, University of Toronto, ON M5S 3H4, Canada}
\affiliation{Dunlap Institute for Astronomy and Astrophysics, 50 St. George Street, University of Toronto, ON M5S 3H4, Canada}

\author[0000-0001-9345-0307]{Victoria M. Kaspi}
\email{victoria.kaspi@mcgill.ca}
\affiliation{Department of Physics, McGill University, 3600 rue University, Montr\'eal, QC H3A 2T8, Canada}
\affiliation{Trottier Space Institute, McGill University, 3550 rue University, Montr\'eal, QC H3A 2A7, Canada}
\affiliation{School of Physics and Astronomy, Tel Aviv University, Tel Aviv 69978, Israel}

\author[0009-0004-4176-0062]{Afrokk Khan}
\email{afrasiyab.khan@mcgill.ca}
\affiliation{Department of Physics, McGill University, 3600 rue University, Montr\'eal, QC H3A 2T8, Canada}
\affiliation{Trottier Space Institute, McGill University, 3550 rue University, Montr\'eal, QC H3A 2A7, Canada}

\author[0009-0008-6166-1095]{Bikash Kharel}
  \email{bk0055@mix.wvu.edu}
  \affiliation{Department of Physics and Astronomy, West Virginia University, PO Box 6315, Morgantown, WV 26506, USA}
  \affiliation{Center for Gravitational Waves and Cosmology, West Virginia University, Chestnut Ridge Research Building, Morgantown, WV 26505, USA}
  
\author[0000-0003-4584-8841]{Lluis Mas-Ribas}
\email{lmasriba@ucsc.edu}
\affiliation{Department of Astronomy and Astrophysics, University of California, Santa Cruz, 1156 High Street, Santa Cruz, CA 95060, USA}

\author[0000-0002-4279-6946]{Kiyoshi W. Masui}
\email{kmasui@mit.edu}
\affiliation{MIT Kavli Institute for Astrophysics and Space Research, Massachusetts Institute of Technology, 77 Massachusetts Ave, Cambridge, MA 02139, USA}
\affiliation{Department of Physics, Massachusetts Institute of Technology, 77 Massachusetts Ave, Cambridge, MA 02139, USA}

\author[0000-0002-0786-7307]{Kyle McGregor}
\email{kyle.mcgregor@mail.mcgill.ca}
\affiliation{Department of Physics, McGill University, 3600 rue University, Montr\'eal, QC H3A 2T8, Canada}
\affiliation{Trottier Space Institute, McGill University, 3550 rue University, Montr\'eal, QC H3A 2A7, Canada}

\author[0000-0002-2551-7554]{Daniele Michilli}
\email{danielemichilli@gmail.com}
\affiliation{Laboratoire d'Astrophysique de Marseille, Aix-Marseille Univ., CNRS, CNES, Marseille, France}

\author[0000-0001-7348-6900]{Ryan Mckinven}
\email{ryan.mckinven@mcgill.ca}
\affiliation{Department of Physics, McGill University, 3600 rue University, Montr\'eal, QC H3A 2T8, Canada}
\affiliation{Trottier Space Institute, McGill University, 3550 rue University, Montr\'eal, QC H3A 2A7, Canada}

\author[0000-0002-0940-6563]{Mason Ng}
\email{mason.ng@mcgill.ca}
\affiliation{Department of Physics, McGill University, 3600 rue University, Montr\'eal, QC H3A 2T8, Canada}
\affiliation{Trottier Space Institute, McGill University, 3550 rue University, Montr\'eal, QC H3A 2A7, Canada}

\author[0000-0003-0510-0740]{Kenzie Nimmo}
\email{knimmo@northwestern.edu}
\altaffiliation{NASA Einstein Fellow.}
\affiliation{Center for Interdisciplinary Exploration and Research in Astronomy, Northwestern University, 1800 Sherman Avenue, Evanston, IL 60201, USA}

\author[0009-0008-7264-1778]{Swarali Shivraj Patil}
  \email{sp00049@mix.wvu.edu}
  \affiliation{Department of Physics and Astronomy, West Virginia University, PO Box 6315, Morgantown, WV 26506, USA}
  \affiliation{Center for Gravitational Waves and Cosmology, West Virginia University, Chestnut Ridge Research Building, Morgantown, WV 26505, USA}
\author[0000-0002-8912-0732]{Aaron~B.~Pearlman}
\email{aaron.b.pearlman@mit.edu}
\altaffiliation{NASA Hubble Fellow.}
\affiliation{MIT Kavli Institute for Astrophysics and Space Research, Massachusetts Institute of Technology, 77 Massachusetts Ave, Cambridge, MA 02139, USA}
\affiliation{Department of Physics, Massachusetts Institute of Technology, 77 Massachusetts Ave, Cambridge, MA 02139, USA}
\affiliation{Department of Physics, McGill University, 3600 rue University, Montr\'eal, QC H3A 2T8, Canada}
\affiliation{Trottier Space Institute, McGill University, 3550 rue University, Montr\'eal, QC H3A 2A7, Canada}

\author[0000-0002-4623-5329]{Mawson W. Sammons}
\email{mawson.sammons@mcgill.ca}
\affiliation{Department of Physics, McGill University, 3600 rue University, Montr\'eal, QC H3A 2T8, Canada}
\affiliation{Trottier Space Institute, McGill University, 3550 rue University, Montr\'eal, QC H3A 2A7, Canada}

\author[0000-0003-3154-3676]{Ketan R. Sand}
\email{ketan.sand@mail.mcgill.ca}
\affiliation{Department of Physics, McGill University, 3600 rue University, Montr\'eal, QC H3A 2T8, Canada}
\affiliation{Trottier Space Institute, McGill University, 3550 rue University, Montr\'eal, QC H3A 2A7, Canada}
\author[0009-0005-8384-320X]{Aylar	Sedaei Oghani}
\email{aylar@yorku.ca}
\affiliation{Department of Physics and Astronomy, York University, 4700 Keele Street, Toronto, ON MJ3 1P3, Canada}

\author[0000-0002-4823-1946	]{Vishwangi Shah}
\email{vishwangi.shah@mail.mcgill.ca} 
\affiliation{Department of Physics, McGill University, 3600 rue University, Montr\'eal, QC H3A 2T8, Canada}
\affiliation{Trottier Space Institute, McGill University, 3550 rue University, Montr\'eal, QC H3A 2A7, Canada}

\author[0000-0002-2088-3125]{Kendrick Smith}
\email{kmsmith@perimeterinstitute.ca}
\affiliation{Perimeter Institute of Theoretical Physics, 31 Caroline Street North, Waterloo, ON N2L 2Y5, Canada}
\author[0000-0001-9784-8670]{Ingrid Stairs}
  \email{stairs@astro.ubc.ca}
  \affiliation{Department of Physics and Astronomy, University of British Columbia, 6224 Agricultural Road, Vancouver, BC V6T 1Z1 Canada}
\author[0000-0002-7076-8643]{Tarik J. Zegmott}
\email{tarik.zegmott@mcgill.ca}
\affiliation{Department of Physics, McGill University, 3600 rue University, Montr\'eal, QC H3A 2T8, Canada}
\affiliation{Trottier Space Institute, McGill University, 3550 rue University, Montr\'eal, QC H3A 2A7, Canada}
\begin{abstract}
We present 30 newly discovered repeating fast radio burst (FRB) sources from the second catalog of bursts detected by the FRB backend on the Canadian Hydrogen Intensity Mapping Experiment (CHIME/FRB). These repeaters have extragalactic dispersion measures (DMs) spanning $99.4-1446.0\ \dmunits$ and burst rates between $10^{-5.7}$  and $10^{-0.5}$ hr$^{-1}$ scaled to a fluence threshold of 5 Jy ms. We report evidence of monotonic, linear DM variations in four repeaters on years-long timescales. The newly discovered sources bring CHIME/FRB's total number of observed repeating FRBs to 80, 79 of which were discovered by CHIME/FRB, between 2018 July 25 and 2023 September 15. In the full CHIME/FRB sample, only 2.4$\pm 0.4\%$ of sources have been observed to repeat, and we do not find evidence for significant evolution of this value over the duration of the experiment. We find no substantial evidence for bimodal populations of one-off and repeating FRBs in their burst rate distributions; the distribution of upper limits on repeat rates implied from observations of as-yet one-offs is entirely contained within the observed range of repeater burst rates and the distributions do not appear inconsistent. Similarly, using the population analysis framework of \cite{2023PASA...40...57J}, we find that our observations of repeating and yet-one-off FRBs are equally well fit assuming a power-law distribution of repeat rates with 50--100\% of the population repeating. 
\end{abstract}


\shorttitle{CHIME/FRB's Fourth Repeater Catalog}

\shortauthors{Cook, A.M., Shin, K., Pleunis, Z. et al.}

\section{Introduction} \label{sec:intro}

The field of fast radio bursts (FRBs) was transformed with the  discovery of multiple bursts coming from a single source \citep{R1}. Prior to this observation, many source theories for FRBs drew connections to cataclysmic events given the large energies implied by the brightnesses ($\sim$ Jy) and inferred distances (extragalactic) of these micro- to milli- second signals \citep[e.g.,][]{2013Sci...341...53T,10.1093/mnrasl/slw069}. The discovery of their repetition ruled out the entire class of cataclysmic source models for these so-called \textit{repeaters}.

The source(s) of FRBs, observed to be repeating or otherwise, remain elusive. The detection of an FRB-like, $\sim$ millisecond duration radio burst from a Galactic magnetar (\citetalias{2020Natur.587...54C} \citeyear{2020Natur.587...54C}; 
\citealt{2020Natur.587...59B}) implies that at least some FRBs can be explained by sporadic magnetar emission \citep{2020ApJ...899L..27M}. The FRB-like Galactic magnetar burst was brighter than the least luminous observed extragalactic FRBs, particularly those observed from the repeating source FRB~20200120E, the nearest known FRB that resides in a globular cluster associated with M81 \citep{2021ApJ...910L..18B,2022Natur.602..585K,2022NatAs...6..393N}. 
This environment seems to be in tension with a magnetar source, however, given that magnetars are typically thought to be the young products of core-collapse supernovae of massive stars, and globular clusters are much too old to host such systems. FRB~20240209A, which was localized to the outskirts of a massive and quiescent host, provides another example of an environment not obviously expected for a magnetar.\citep{2025ApJ...979L..21S,Eftekhari_2025}. 

 Repeaters are the most promising sources for detecting burst-like multi-wavelength counterparts to FRBs given that we can coordinate pointed, high-angular-resolution, and sensitive observations at the time of known radio activity for repeating FRB sources, and we have yet to detect such a counterpart \citep{2016ApJ...833..177S,2020ApJ...901..165S,2024ApJ...974..170C,2024ApJ...964..121K,2025A&A...695L..10E,2025NatAs...9..111P}. 

Repeater observations allow us to track not only the evolution of the sources through properties like burst rates, but also the evolution of their local environments through repeated measurements of the dispersion measure (DM), scattering and scintillation, and Faraday rotation measure (RM) enabled by the multiple bursts \citep{2018Natur.553..182M,Hilmarsson_2021,2022NatCo..13.4382W,2023MNRAS.526.3652K,2023ApJ...950...12M,2023ApJ...951...82M,10.1093/mnras/stac3547,2025ApJ...982..154N,alice,2025arXiv251011352S,NIU202676}. Additionally, repeaters give us an insight into the total energy budget of the sources \citep{2025arXiv250714707Z,2024arXiv241017024O,thomasr117}. 

Repeater burst arrival times are stochastic; however, they cluster in time with waiting time distributions well described by the Weibull distribution with sub-Poissonian shape parameter \citep{2016ApJ...833..177S,2018MNRAS.475.5109O,2020A&A...635A..61O,2022Natur.606..873N}. In some cases, these activity periods are periodic (\citetalias{R3periodicity} \citeyear{R3periodicity};  \citealt{10.1093/mnras/staa1237,10.1093/mnras/staa3223}), e.g., FRB~20180916B bursts stochastically with a duty cycle of $\sim 5$ days on a $16.34\pm 0.07$ day period \citep{2023ApJ...956...23S}. The  periodic activity windows of FRB~20180916B are also chromatic, occurring later for lower frequencies \citep{2021ApJ...911L...3P,2021Natur.596..505P,2023MNRAS.524.3303B,2025arXiv250704609E}.

A question of great interest is `Do all FRBs repeat?', and observational evidence has yet to unambiguously rule either way. Indeed, no FRB has been definitively associated with a cataclysmic event \citep[e.g., ][]{2023ApJ...955..155A,2023NatAs...7..579M,2023ApJ...954..154C,2024ApJ...972..125C,2025ApJ...991..199D,2024MNRAS.52711158Y}. \cite{RN3} found no evidence for bimodality in implied burst rates between the population of repeating and upper limits from as-yet non-repeating FRBs. \cite{2024NatAs...8..337K} and \cite{2024arXiv241017024O} found that a single source luminosity distribution can explain the population of apparent one offs and repeaters given that the luminosity distribution flattens out at high energies, albeit such a luminosity distribution has only been measured for a few, hyperactive, sources that may not be representative of all repeating sources. Population-level modeling similarly can reconcile the two populations \citep[e.g.,][]{frbpoppy,2023PASA...40...57J,2024ApJ...961...10M,2025ApJ...993...37B} and there has been no evidence that the local environments, as far as they are probed by scattering timescales and RMs, are statistically inconsistent between repeaters and apparent one-offs  \citep{2024ApJ...968...50P,alice}.

Phenomenologically, however, there is a significant difference between the two populations, with the repeater population typically having narrower bandwidth and longer duration on average compared to the apparent non-repeating population \citep{pleunismorph,alice}. The different phenomenology may be indicative of different source classes. In order to reconcile these observations with a single population/emission mechanism, the activity rates must be somehow correlated and anti-correlated with duration and bandwidth, respectively \citep{pleunismorph}. Such correlations could arise either intrinsically from the emission mechanism, extrinsically (e.g., related to a propagation effect due to differing environments), or through a selection effect \citep[see, for example, ][]{2020MNRAS.497.3076C}.

The wide field of view, regular daily observing cadence, and high sensitivity of the FRB backend of the Canadian Hydrogen Intensity Mapping Experiment (CHIME/FRB) make it a particularly powerful repeater detector \citep{2018ApJ...861L...1C}, with CHIME/FRB having discovered the vast majority of published repeaters to date (\citealt{R1}; \citetalias{R2} \citeyear{R2}; \citealt{2019ApJ...887L..30K}; \citetalias{RN1} \citeyear{RN1}; \citealt{2020Natur.586..693L,RN2,2021MNRAS.500.2525K,2022ATel15679....1M}; \citetalias{RN3} \citeyear{RN3}; \citealt{2024ATel16734....1A,2025ApJ...979L..21S,2025arXiv250513297S, 2025MNRAS.540.1685T}). In this paper, we present the uniform population statistics of \ntotalrepeaters{} repeaters detected by CHIME/FRB, 79 of which were discovered by CHIME/FRB, and including  \nrepeaters{} newly discovered sources identified in CHIME/FRB's Second FRB catalog (\citetalias{cat2} \citeyear{cat2})  and that we report here for the first time. In \S \ref{section:CatalogII}, we give a brief overview of the CHIME/FRB survey and detail our process for identifying new repeater candidates and quantifying their  significance. For those repeating FRB candidate sources which meet our significance criteria, we present updated source parameters including position and burst rate, and examine the repeater population in \S \ref{sec:repprop}. In \S\ref{sec:r_v_o}, we compare the repeating sample from CHIME/FRB to the apparently non-repeating sample. Finally, in \S\ref{sec:disc}, we discuss implications for the population and the notable sources in the sample.

\section{Observations} 
\label{section:CatalogII}
\subsection{CHIME/FRB and Catalog 2}

The CHIME telescope is a transit radio telescope that operates across the 400--800 MHz frequency range and has a large instantaneous field of view \cite[$\sim 200$ deg$^2$; ][]{2022ApJS..261...29C}. The CHIME/FRB backend searches beam-formed total intensity data from CHIME's FX correlator in real time for FRBs and is described by \cite{chimefrboverview}. The sample from which we identify repeaters is CHIME/FRB's Second catalog, henceforth \catii{} (\citetalias{cat2} \citeyear{cat2}). Briefly, this sample contains the \nfrbcat{} bursts detected by CHIME/FRB between 25 July 2018 and 15 September 2023 that meet all criteria for inclusion. Where available, we use the increased burst localization precision afforded by the offline `beamforming' raw-voltage data (more detail about the localizations can be found in \S\ref{sec:localization}). The criteria for inclusion in Cat2, listed in detail in \S 2 of \citetalias{cat2} (\citeyear{cat2}), include having measured burst DMs above both the estimates of the maximum line-of-sight DM contribution from the ISM of our Milky Way from \cite{ne2001} and \cite{ymw16} (NE2001 and YMW16, respectively), and positive verification as a FRB candidate by at least two experts. 

This work represents the fourth catalog of repeating sources from CHIME/FRB, and where analysis remains unchanged from previous catalogs we explicitly refer the reader to those works, which are RepCat1 (\citetalias{RN1} \citeyear{RN1}), RepCat2 \citep{RN2}, and RepCat3 (\citetalias{RN3} \citeyear{RN3}).  

\subsection{Source Identification}
\label{section:SourceIdentification}
In order to identify new repeaters in the CHIME/FRB sample of FRBs, we first identify groups of bursts that are close to each other in R.A., Decl., and DM using a clustering algorithm. This returns order 100 clusters, including many that may be consistent with each burst coming from a distinct source, and the sources are proximal to one another by chance. The ambiguity primarily arises from CHIME/FRB's non-negligible localization uncertainties compared to our detection density on the sky. We thus diagnose the significance of a cluster by computing the probability of detecting that many independent FRB sources within the cluster's volume as the probability of chance coincidence, or \pcc{}. In order to compute this probability, we need some description of CHIME/FRB's detection rate in R.A., Decl., and DM. We use the noisy, non-homogeneous Poisson process intensity function estimated by \cite{Cook2024KContact} for CHIME/FRB for this purpose. The intensity function is a hierarchical Bayesian fit of CHIME/FRB data to a parameterization that accounts for CHIME/FRB's exposure, sensitivity, spherical geometry, non-negligible and multi-modal localization uncertainties, and the underlying luminosity function of FRBs. We describe these analyses in more detail and the resultant clusters and \pcc{} estimates in the following subsections. 
 
\subsubsection{Clustering Algorithm}
\label{sec:clustering}
Candidate repeaters, or clusters, are identified via a DBSCAN-based \citep{10.5555/3001460.3001507} clustering algorithm \citep[described by ][]{2023MNRAS.524.5132D} run on all bursts in \catii{}, considering their positions and DMs, as in RepCat3. The nearest neighbour distance, $\epsilon$ is based on our uncertainty in R.A., Decl., and DM. The uncertainty in R.A. for a given event is highly dependent on the N--S zenith angle and hence is set to 2.2~degrees~$/ \cos(\text{Decl.})$. We assume constant fiducial values for this $\epsilon$ value in the clustering algorithm of 0.5 degrees and 13 $\dmunits{}$ and in Decl. and DM, respectively. 

 Temporal DM variation for a single source on this scale has been observed from repeaters, especially when considering the multi-year timescales like that of our sample  
 \citep[$\sim 0.1 -12\,  \dmunits{} \text{ yr}^{-1}$;][]{2023MNRAS.526.3652K,2025arXiv250715790W,2025arXiv251011352S,2025arXiv250916374O,NIU202676,pandhi26,thomasr117}. Depending on how many bursts we have observed from the source and the scale of the DM variation, the clustering algorithm may not identify such sources as repeaters. We further discuss our search for variable DM sources in \S \ref{sec:deltadm}.

\subsubsection{Chance Coincidence Probability Calculation}

In its candidate identification, the clustering algorithm does not account for the nonhomogeneous nature of CHIME/FRB's exposure and sensitivity as a function of sky position and DM. This non-homogeneity arises primarily from the instrument's declination-dependent exposure and zenith-angle-dependent sensitivity. We thus must assess the significance for each of the \nclusters\, candidate clusters identified by the algorithm. For this significance, we want to estimate the \pcc{}, that we define as the probability of CHIME/FRB observing $k$ separate FRB sources in that region of R.A., Decl., DM space in \catii{} by chance, where $k$ is the number of FRBs in a candidate cluster. 

To do so, we use the methodology proposed and applied to the CHIME/FRB survey by \cite{Cook2024KContact}. This methodology allows for the estimation of the $k$-contact distance for events in a nonhomogeneous Poisson process. The $k$-contact distance distribution estimates, for some position in our R.A., Decl., DM observation space, how likely it is to have observed $k$ or more events within a given radius. The calculation of the $k$-contact distance assumes an intensity function\footnote{The intensity function can be thought of as a spatially dependent rate for a Poisson process in multiple dimensions.}, which \cite{Cook2024KContact} estimated for a physically motivated parameterization of the survey fit using CHIME/FRB's First catalog of FRB sources (henceforth \cati; \citealt{first_chime_catalog2021ApJS..257...59C}). 

The main improvements in this methodology, compared to that employed by \cite{RN3} for RepCat3, are a full uncertainty quantification in the model inferences and a physically motivated intensity parameterization. 
In RepCat3, the intensity function was estimated by smoothing a two dimensional (DM and $\delta$) detection histogram with a Gaussian kernel, and the probability of detection is approximated with binomial statistics as the fraction of the detection probability density inside/outside of the box. An ambiguity arises in this previous methodology when defining the `in' box. The previous methodology also does not take into account measurement error other than the 95\% span of the localization and DM errors, inside of which positions and DMs are equally weighted despite the highly multi-modal nature of the header localizations. The arbitrary choice of a Gaussian kernel could potentially obscure minute fluctuations in intensity, or could overweight a stochastic over- or under- density. 

The hierarchical Bayesian approach to model fitting of \cite{Cook2024KContact} also allows for uncertainty quantification. \cite{Cook2024KContact} define an intensity parameterization based on the properties of the CHIME/FRB survey and set priors informed by independent observations. We sample from the localization uncertainty regions to define the minimal bounding sphere (which defines the radius tested in the $k$-contact distance) and sample from the posterior distribution of the intensity function when computing the final \pcc{}  statistic (Eqn. 7 from \citealt{Cook2024KContact}). This allows for estimates of uncertainty for the \pcc{} values. 

The probability of detecting $k$ separate FRB sources within some radius depends on the total number of FRBs detected and how many trials we perform. The former is accounted for by integrated intensity over the observation domain. To account for the additional FRBs detected in Cat2 compared to Cat1, we scale the $N_{\text{FRBs}}$/normalization parameter posterior (which was based on the 492 Cat1 sources and has a posterior median estimate of $493^{+23}_{-22}$ FRBs) by a multiplicative factor of $N_{\text{Cat2}}/N_{\text{Cat1}}$, or 3641/492. The choice of trials factor is discussed in more detail in the next section.

\begin{figure*}
    \centering
    \includegraphics[width=\linewidth]{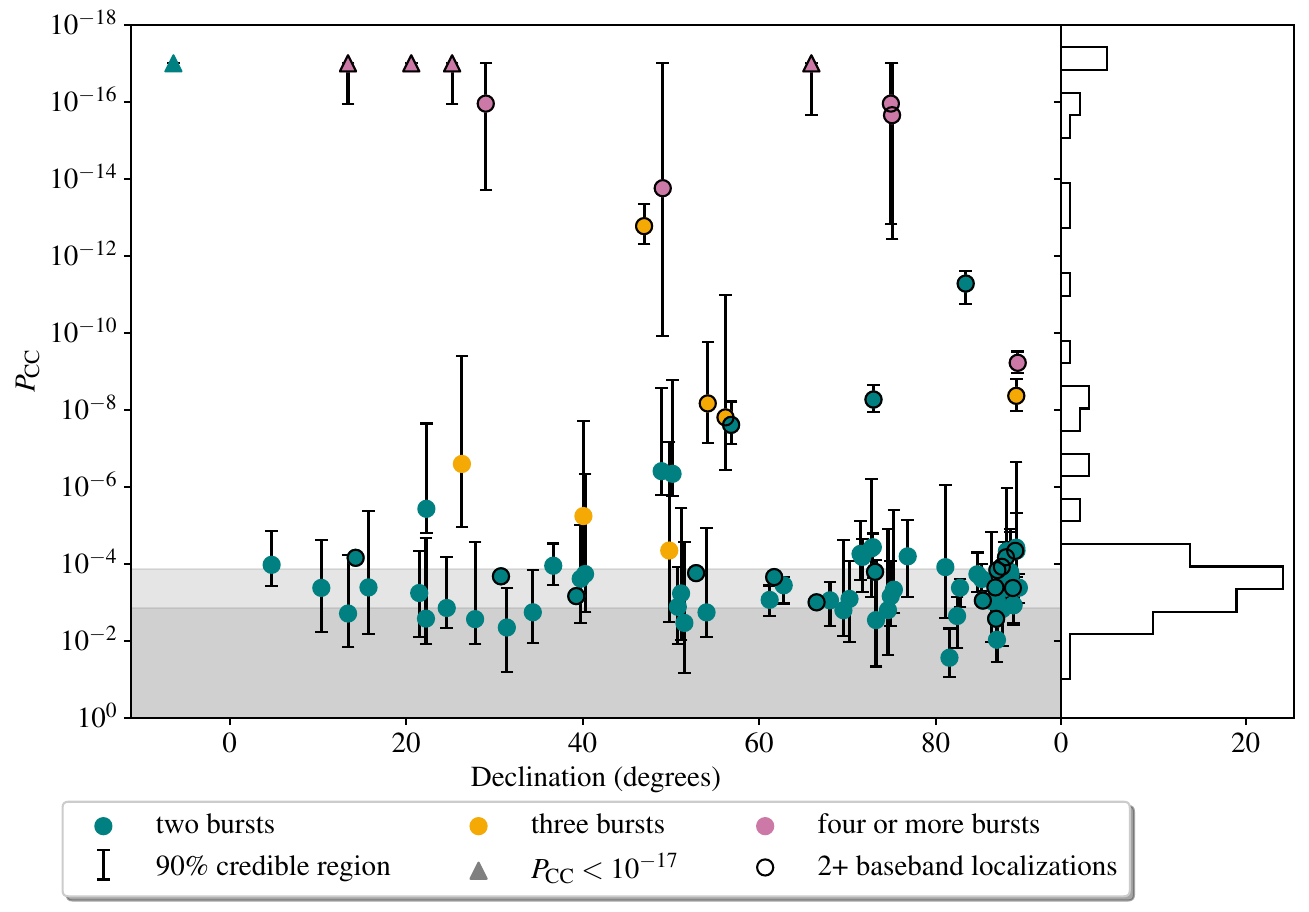}
    \caption{Probability of chance coincidence (\pcc) versus declination of repeater candidates from CHIME/FRB's Second catalog. Candidates are colored according to the number of bursts in their clusters, which were identified via an in-house implementation of a DBSCAN clustering algorithm (see \S \ref{sec:clustering}). The error bars on the data points  correspond to the candidates' 95\% credible regions, where uncertainties have arisen due to the uncertainty in localization and the uncertainty in the posterior estimate of our model of CHIME/FRB's detections \citep{Cook2024KContact}. The unshaded region corresponds to median \pcc{} values in our `gold' sample, the lighter gray region corresponds to `silver' or candidate FRB, and the darker gray region corresponds to median \pcc{} values considered indistinguishable from a chance coincidence.  These \pcc{} values decrease considerably with improved localization precision, and hence the clusters with multiple baseband-localized bursts are outlined in black. The panel on the right shows a histogram of the medians of these \pcc\, estimates. Our double precision floats cannot meaningfully distinguish \pcc{} values below $10^{-17}$. For sources in this regime we therefore report only upper limits on \pcc{}. These upper limits are shown with triangle markers. The hatched region in the marginal plot also corresponds to upper-limit values of \pcc{}.}
    \label{fig:pcc}
\end{figure*}

The \pcc{} itself has some associated uncertainty, given we sample from the (scaled) posterior distribution of all unknown model parameters of the intensity. The median estimated \pcc{} for all repeater candidates from Cat2, and error bars corresponding to the width of the 90\% credible region\footnote{Specifically the highest posterior density 90\% credible region.}, are shown in Figure \ref{fig:pcc}. Our double precision floats cannot meaningfully distinguish \pcc{} values below $10^{-17}$. For sources in this regime we therefore report only upper limits on \pcc{}.

Our threshold for the gold sample is adopted from RepCat3 and are applied to the median \pcc{} estimates, but we additionally enforce a constraint on the 90\% credible region. That is, we consider sources to have met the threshold for inclusion in the gold sample when both the median \pcc{} $< 0.5/N_{\text{trials}} = 0.5/3641$ (as in RepCat3) and the entirety of the 90\% credible region of the \pcc{} is $< 1/N_{\text{trials}} = 1/3641$. 
\nrepeaters{} of the 93 input clusters meet the criteria for the gold sample. These sources and their properties are listed in Table \ref{tab:rn4prop}. 

 Sources in the silver sample are those which are not in the gold sample, have median \pcc{} $< 1/3641$, and the entirety of the 90\% credible region of their \pcc{} is $<5/3641$.  A total of \nsilver{} FRB sources matched the criteria for the silver sample. We consider these sources only candidate repeaters, and report them and their properties in Appendix Table \ref{tab:silverprop}.

\subsubsection{Controlling the Family-Wise Error Rate (selecting a trials factor) }

 The most accurate trials factor would reflect the DBSCAN search that CHIME/FRB uses to identify the tens to hundreds of repeater candidate clusters, but it cannot be represented as a single grid search given the $\epsilon$-neighborhood for each event in our clustering algorithm is based on its declination and positional uncertainty. Relatedly, it is not correct to choose an area of the smallest region tested and divide by the total surveyed area, since the probability of detection varies as a function of position on the sky. Instead, we use the number of detected FRBs for the trials factor, which is conservative as the largest sensible choice: it would be the appropriate trials factor for a search where we choose a radius around each FRB which maximizes the density and then compute this $k$-contact distance probability for that radius. In reality, we are only testing the \pcc{} for a small subset of FRBs, but it is reasonable to assume the minimal bounding sphere of those chosen clusters maximizes the density around the few we do search to some extent. 
\subsection{Updated repeater status ($P_{\text{CC}}$) for previous silver sources}
\label{sec:silverpcc}
 RepCat3, which was based on observations from a shorter time range than this work, reported 14 additional repeater candidates, referred to as the `silver' sample sources, which were promising clusters of bursts with \pcc{} at an ambiguous level. While we expect the increase in total number of FRBs detected to increase the \pcc, and so decrease the significance of a given candidate, the new \pcc{} methodology has been demonstrated to be more sensitive. Furthermore, additional bursts may have been detected in the \catii{} period, decreasing the \pcc. Thus we revisit the \pcc{} of these candidates in this work given the \catii{} sample. First, we consider any additional bursts from \catii{} which were grouped with these repeaters by the clustering algorithm. Three of the RepCat3 silver sources, FRB~20181201D, FRB~20190905A, and FRB~20230323C had two, four, and three additional bursts, respectively, detected after the cut off of RepCat3. Next, we compute the updated \pcc{} probabilities according to the new methodology and using the appropriate \catii{} trials factor of \nsources{}. The updated \pcc{} values are listed in Table \ref{tab:silverpcc}.

 All three RepCat3 silver candidates for which we detected additional candidates now pass the threshold for a gold-sample repeater.  We include these three sources and their properties in Table \ref{tab:rn4prop}, but note they were originally presented in RepCat3.  Five of the candidates remain candidates, with updated \pcc{} in our current silver threshold. An additional five RepCat3 candidates are deemed indistinguishable from random chance alignment of two or more sources given our localization precision and the Cat2 data. One RepCat3 candidate source, FRB~20200508H, had only one burst present in Cat2, the other associated burst did not meet the criteria for inclusion for Cat2 owing to inadvertent deletion of the total intensity data and so we do not report an updated \pcc{} in this work (see RepCat3 for more details).

\startlongtable
\begin{longrotatetable}
\begin{deluxetable}{lllllllllll}
\tablehead{
\colhead{Source} & \colhead{$P_{\text{CC}}$} & \colhead{R.A.\tablenotemark{a}} & \colhead{Decl.\tablenotemark{a}}  & \colhead{DM} & \colhead{$\text{DM}_{\text{NE2001}}$} & \colhead{$\text{DM}_{\text{YMW16}}$} & \colhead{$N$\tablenotemark{b}} & \colhead{Exposure} & \colhead{Sensitivity\tablenotemark{c}} & \colhead{Rate $ \geq \text{5 Jy ms}$\tablenotemark{d}}\\ 
\colhead{} & \colhead{} & \colhead{(deg)} & \colhead{(deg)}  & \colhead{(pc cm$^{-3}$)} & \colhead{(pc cm$^{-3}$)} & \colhead{(pc cm$^{-3}$)} & \colhead{} & \colhead{(U/L, hr)} & \colhead{(U/L, Jy ms)} & \colhead{(hr$^{-1}$)}}
\tablecaption{Properties of New Repeating Sources of FRBs, 
(Our “Gold” Sample)\label{tab:rn4prop}.}
\startdata
FRB~20220529A\tablenotemark{e} & $\lesssim 10^{-17}$ & 19.10422(8) & 20.632(5) & 245.698(5) & 214.8 & 205.7 & 15 & 90(10) & 5.3 & $7_{-5}^{+7}\times 10^{-2}$ \\
FRB~20210601A & $\lesssim 10^{-17}$ & 200.782(7) & 65.837(6) & 418.94(2) & 392.7 & 384.8 & 9 & 220(10) & 0.2 & $0.3_{-0.2}^{+1.0}\times 10^{-4}$ \\
FRB~20210822E\tablenotemark{f} & $\lesssim 10^{-17}$ & 216.556(5) & 25.244(5) & 319.452(5) & 298.1 & 298.1 & 6 & 108(2) & 9.0 & $1.5_{-0.8}^{+5.0}\times 10^{-2}$ \\
FRB~20220618C\tablenotemark{f} & $\lesssim 10^{-17}$ & 46.218(5) & 13.405(5) & 187.717(6) & 146.5 & 141.4 & 4 & 111.1(1) & 10.5 & $1.8_{-0.9}^{+6.0}\times 10^{-2}$ \\
FRB~20191026A & $\lesssim 10^{-17}$ & 55.6(2) & -6.4(5) & 147.8(1) & 103.8 & 108.4 & 2 & 60(40) & 24.8 & $0.3_{-0.3}^{+2.0}$ \\
FRB~20190929E & $1\times 10^{-16}$ & 311.15(2) & 28.92(2) & 513.03(6) & 397.6 & 352.3 & 7 & 80(10) & 8.6 & $1.1_{-0.9}^{+1.0}\times 10^{-1}$ \\
FRB~20220505A & $1\times 10^{-16}$ & 29.980(5) & 74.982(5) & 605.923(7) & 456.5 & 496.7 & 8 & 350(20)/327(1) & 3.8/20.2 & $5_{-4}^{+7}\times 10^{-3}$ \\
FRB~20210323C\tablenotemark{g} & $1\times 10^{-16}$ & 122.199(8) & 72.431(8) & 287.935(8) & 242.6 & 237.4 & 6 & 200(100)/200(100) & 0.2/1.8 & $1_{-1}^{+5}\times 10^{-4}$ \\
FRB~20210409B\tablenotemark{f}& $2\times 10^{-16}$ & 292.522(7) & 75.162(7) & 692.121(4) & 633.8 & 632.3 & 7 & 442(4)/290(10) & 23.4/4.8 & $1.4_{-0.7}^{+5.0}\times 10^{-2}$ \\
FRB~20190905A\tablenotemark{g} & $3\times 10^{-15}$ & 51.105(5) & 89.583(6) & 232.831(2) & 179.6 & 177.2 & 6 & 3800(1300)/3400(1000) & 14.6/10.2 & $2_{-2}^{+6}\times 10^{-3}$ \\
FRB~20210209C & $2\times 10^{-14}$ & 203.424(5) & 49.088(5) & 382.40(2) & 360.8 & 353.4 & 5 & 163(2) & 0.2 & $2_{-1}^{+2}\times 10^{-4}$ \\
FRB~20210125A & $2\times 10^{-13}$ & 338.575(9) & 46.951(9) & 459.42(3) & 329.4 & 322.2 & 3 & 155(5) & 3.4 & $1_{-1}^{+2}\times 10^{-2}$ \\
FRB~20190208C & $5\times 10^{-12}$ & 126.071(4) & 83.400(4) & 238.3167(6) & 190.7 & 186.5 & 2 & 640(30)/590(40) & 0.10/0.2 & $0.3_{-0.2}^{+1.0}\times 10^{-5}$ \\
FRB~20190224D & $6\times 10^{-10}$ & 245.801(7) & 89.081(5) & 751.903(3) & 701.2 & 697.9 & 4 & 4700(200)/6900(100) & 3.0/0.2 & $0.6_{-0.3}^{+2.0}\times 10^{-4}$ \\
FRB~20181201D\tablenotemark{g} & $1\times 10^{-9}$ & 235.57(1) & 88.523(5) & 445.052(2) & 395.9 & 392.1 & 4 & 6300(2100)/7200(1700) & 2.0/2.3 & $0.3_{-0.2}^{+1.0}\times 10^{-4}$ \\
FRB~20190320E & $4\times 10^{-9}$ & 78.221(5) & 89.114(6) & 308.786(2) & 255.3 & 253.0 & 3 & 7300(300)/7370(30) & 3.3/2.9 & $1_{-1}^{+2}\times 10^{-4}$ \\
FRB~20210330B & $5\times 10^{-9}$ & 80.023(6) & 72.953(6) & 397.38(1) & 309.3 & 321.1 & 2 & 260(20)/299(1) & 0.04/0.3 & $2_{-1}^{+8}\times 10^{-6}$ \\
FRB~20220319C & $7\times 10^{-9}$ & 223.804(5) & 54.225(5) & 307.58(4) & 282.9 & 275.3 & 3 & 30(20) & 12.9 & $0.1_{-0.1}^{+1.0}$ \\
FRB~20210924C & $2\times 10^{-8}$ & 309.02(1) & 56.15(1) & 494.58(3) & 299.2 & 351.2 & 3 & 180(10) & 2.2 & $4_{-4}^{+6}\times 10^{-3}$ \\
FRB~20210504D & $2\times 10^{-8}$ & 201.448(4) & 56.820(4) & 564.560(4) & 541.4 & 533.4 & 2 & 206.7(5) & 2.0 & $3_{-3}^{+5}\times 10^{-3}$ \\
FRB~20220101A & $3\times 10^{-7}$ & 245.25(1) & 26.41(1) & 385.78(4) & 356.2 & 352.7 & 3 & 67(13) & 1.6 & $4_{-4}^{+9}\times 10^{-3}$ \\
FRB~20220123A\tablenotemark{f} & $4\times 10^{-7}$ & 202.449(5) & 48.848(5) & 120.857(2) & 99.4 & 92.1 & 2 & 130.4(1) & 0.9 & $0.4_{-0.2}^{+1.0}\times 10^{-3}$ \\
FRB~20220428C & $5\times 10^{-7}$ & 130.928(8) & 50.191(8) & 140.038(3) & 102.4 & 94.2 & 2 & 182(1) & 5.7 & $2_{-1}^{+2}\times 10^{-2}$ \\
FRB~20200308B & $4\times 10^{-6}$ & 272.029(7) & 22.219(7) & 218.01(1) & 149.9 & 133.8 & 2 & 80(7) & 5.0 & $2_{-2}^{+4}\times 10^{-2}$ \\
FRB~20191113C & $6\times 10^{-6}$ & 137.51(2) & 40.13(2) & 617.95(10) & 585.9 & 574.9 & 3 & 137(5) & 6.6 & $2_{-2}^{+4}\times 10^{-2}$ \\
FRB~20220207A & $4\times 10^{-5}$ & 310.151(5) & 72.909(5) & 262.065(2) & 179.0 & 186.1 & 2 & 230(30)/297(2) & 3.6/23.6 & $2_{-1}^{+6}\times 10^{-3}$ \\
FRB~20200220E & $4\times 10^{-5}$ & 351.424(9) & 89.044(9) & 1500.599(5) & 1446.0 & 1444.0 & 2 & 8400(200)/7000(200) & 4.3/4.2 & $1_{-1}^{+3}\times 10^{-4}$ \\
FRB~20180907D & $5\times 10^{-5}$ & 228.544(6) & 89.086(6) & 1437.555(6) & 1387.3 & 1383.9 & 2 & 4400(200)/6500(200) & 1.6/1.5 & $3_{-2}^{+9}\times 10^{-5}$ \\
FRB~20190716C & $5\times 10^{-5}$ & 281.455(6) & 88.883(6) & 330.47(3) & 278.5 & 275.7 & 2 & 5200(200)/3200(200) & 0.5/3.2 & $0.4_{-0.2}^{+1.0}\times 10^{-5}$ \\
FRB~20220211A\tablenotemark{f} & $5\times 10^{-5}$ & 266.340(7) & 71.382(7) & 274.109(7) & 232.0 & 226.8 & 2 & 377.4(2)/278(2) & 1.4/12.7 & $3_{-1}^{+10}\times 10^{-4}$ \\
FRB~20211030A & $7\times 10^{-5}$ & 189.293(6) & 87.668(5) & 333.5691(6) & 286.8 & 282.3 & 2 & 600(40)/2580(10) & 5.6/1.8 & $1.2_{-0.7}^{+4.0}\times 10^{-3}$ \\
FRB~20211120C & $7\times 10^{-5}$ & 40.924(7) & 14.315(7) & 620.685(6) & 584.1 & 577.1 & 2 & 20(10) & 2.7 & $1_{-1}^{+9}\times 10^{-2}$ \\
FRB~20211003B & $1\times 10^{-4}$ & 180.179(4) & 87.667(4) & 544.2489(9) & 497.3 & 492.9 & 2 & 570(20)/2582(5) & 1.5/0.7 & $2_{-1}^{+7}\times 10^{-4}$ \\
\enddata
\tablenotetext{a}{J2000, ICRS}
\tablenotetext{b}{The total number of bursts detected CHIME/FRB by the source. Includes the bursts detected outside of our FWHM at 600 MHz.}
\tablenotetext{c}{Fluence threshold for the source at the 95\% confidence level.}
\tablenotetext{d}{Burst rate corresponding to the upper transit, and scaled to a fluence threshold of 5 Jy ms from the sensitivity in the previous column assuming a powerlaw index of $\alpha = -1.5$.}
\tablenotetext{e}{Position reported by \cite{2025arXiv250304727L}.}
\tablenotetext{f}{The best-known sky position of this source falls in between the FWHMs at 600 MHz of the synthesized beams. We report the exposure at the beam center of the nearest beam and have scaled the sensitivity accordingly.}
\tablenotetext{g}{Originally reported by \cite{RN3} as a repeater `candidate', but now meets the threshold for inclusion as a repeater, see \S\ref{sec:silverpcc}.}

\end{deluxetable}
\end{longrotatetable}

\begin{deluxetable*}{lcccc}
\tablecaption{\label{tab:silverpcc} Updated \pcc{} of silver sources from RepCat3.}
\tablehead{
\colhead{Source Name} & \colhead{$\log_{10}($\pcc{}$)$\tablenotemark{a}} & \colhead{$\log_{10}($\pcc{}$)$} & \colhead{Status} \\
 & \colhead{(RepCat3)} & \colhead{(this work)} & }
\startdata
FRB~20190303D & $-3.55$ & $-2.39$ & $-$\\
FRB~20181201D & $-3.54$ & $-9.02$ & repeater\\
FRB~20190328C & $-3.49$ & $-3.07$ & candidate \\
FRB~20191105B & $-3.48$ & $-3.30$ & candidate\\
FRB~20190107B & $-3.45$ & $-3.35$ & candidate\\
FRB~20200320A & $-3.38$ & $-2.33$ & $-$\\
FRB~20190210C & $-3.34$ & $-3.16$ & candidate\\
FRB~20190812A & $-3.20$ & $-3.55$ & candidate\\ 
FRB~20200828A & $-3.14$ & $-2.66$ & candidate\\
FRB~20190905A & $-3.00$ & $-14.56$& repeater \\
FRB~20190127B & $-2.91$ & $-2.41$ & candidate\\
FRB~20210323C & $-2.86$ & $-15.95$& repeater\\
FRB~20180909A & $-2.75$ & $-2.71$ & candidate\\
\enddata
\tablenotetext{a}{These \pcc{} have been modified from their original `$R_{\text{CC}}$' values (rate of chance coincidence) by dividing by their trials factor (2196) for a more direct comparison.}
\end{deluxetable*}

\section{Repeater properties and population}
\label{sec:repprop}
Individual burst properties are reported in Cat2. For each burst, the total-intensity dynamic spectra are available at 0.98304~ms resolution, as well as parameters derived from morphological burst fits from \texttt{fitburst} \citep{2024ApJS..271...49F}, such as arrival time, intrinsic temporal width, DM, scattering time, and flux density. All of the bursts that we associate with repeaters reported in this work are marked with the corresponding repeater name in Cat2\footnote{The repeater names of all published CHIME/FRB discoveries, as well as FRB~20121102A and FRB~20171019A (originally discovered by \citealt{R1} and \citealt{2019ApJ...887L..30K}, respectively, but which have been detected by CHIME/FRB) are additionally labeled with their repeater name in Cat2.}. In this work, we supplement the per-burst property measurements with global source properties that exploit multiple measurements, namely updated localizations, sensitivity estimations, exposures, measured burst rate, and average DM. We detail these calculations below. 

\subsection{Localization} \label{sec:localization}
For all but one new gold repeater, FRB~20191026A, we have a raw voltage localization of the new sources. This allows us to improve the localization precision substantially by offline `beamforming', which maximizes the telescope sensitivity in any direction within the primary beam \citep[see][]{2021ApJ...910..147M, basetcati}. We provide these updated baseband localizations for the sources in Table \ref{tab:rn4prop},  combining multiple localizations where appropriate to reduce the statistical uncertainty \citep[but retaining the systematic uncertainty, see ][] {2021ApJ...910..147M}. For the other new gold repeater, FRB~20191026A, we do not have an available baseband localization, but exploit our multiple measurements to derive a slightly improved localization for the source. All but three silver sources (FRB~20210305A, FRB~20181018B, and FRB~20200905B) have at least one baseband localization, and for those three we additionally combine the `header' localizations, reported in Table \ref{tab:silverprop}. These localizations are based on per-beam detection S/N and CHIME/FRB's beam model.
These localization are described in more detail by \cite{RN1}, and the authors detail our method to combine multiple header localizations. 

The raw voltage data also allow more precise estimation of burst parameters through their increased frequency and temporal resolution, and since the more precise localization allows us to correct for the beam response of the instrument. Those updated morphological best-fit parameters, individual burst localizations, and flux estimations will be available as a part of our upcoming Second catalog of baseband bursts (CHIME/FRB Collaboration et al. in prep.).

\subsection{Exposure and Sensitivity thresholds}
\label{sec:exposure}
We calculate the total exposure for each source using its refined sky position derived from CHIME/FRB baseband localization (see $\S$\ref{sec:localization}). We follow the CHIME/FRB exposure formalism as described by \cite{first_chime_catalog2021ApJS..257...59C}. For each source, exposure is defined as the cumulative time during which the source is within the full width at half maximum (FWHM) of a CHIME/FRB synthesized detection beam at a reference frequency of 600 MHz and during which the CHIME/FRB system is operating as expected. Exposure for this paper is computed over the interval from 2018 August 28 to 2023 September 15 and is reported in Table \ref{tab:rn4prop}. 

We use a Monte Carlo (MC) simulation to determine sensitivity thresholds, detailed by \citet{2019ApJ...882L..18J}. We account for three main sources of variability that affect burst detectability: i) day-to-day instrument gain variations, ii) changes in synthesized beam response during source transit, and iii) varying emission bandwidths and frequency centers within the instrument band-pass. 
For each instance of the simulation, we draw from a Gaussian distribution centered at the measured fluence of the detected burst, with a width set by its measurement uncertainty. An initial fluence threshold is inferred by scaling the system’s nominal S/N detection threshold by a fluence-to-S/N conversion factor, defined by the trial fluence and the burst's detection S/N. We simulate a random detection scenario to probe the three sources of variability and to estimate a sensitivity scale factor for the simulated FRB compared to the detected burst. The sensitivity scale factor is applied to the initial fluence threshold. The final quoted sensitivity thresholds are defined as the 95th percentile of the distribution of fluence thresholds obtained from the MC realizations.
For repeating sources, to account for intra-source burst variability, each realization of the simulation uses a randomly selected burst from the source as the reference for determining the relative sensitivity and initial fluence threshold. The increased localization precision for these repeating sources compared to Cat2 leads to more precise exposure and fluence estimates and hence an increased precision in our sensitivity thresholds.

\subsection{Burst rate}
\label{sec:rates}
Following the method of previous works (e.g., \citetalias{RN1} \citeyear{RN1}; \citealt{RN2}; \citetalias{RN3} \citeyear{RN3}), we report burst rates referenced to a single fluence threshold of 5 Jy ms by assuming that the cumulative burst rate above a given spectral energy $R(\geq E)$ scales with $E^{-1.5}$. Scaling to a single threshold allows us to more meaningfully compare sources from our instrument, which has a highly declination-dependent sensitivity. A spectral energy cumulative burst rate distribution described by a power-law with index $-1.5$ seems to be a reasonable approximation for moderate spectral energies ($E \lesssim 10^{31}$ erg Hz$^{-1}$, typical measured power-law indices ranging from $-4.6$ to $-1.6$; \citealt{2022ApJ...927...59L,2023ApJ...955..142Z,2024NatAs...8..337K,2024MNRAS.534.3331K,2025MNRAS.540.1685T,2024arXiv241017024O}), and is only observed to be significantly flatter for the highest energy bursts from some repeaters \citep{2023ApJ...955..142Z,2024NatAs...8..337K,2024MNRAS.534.3331K,2024arXiv241017024O}.

The exposures and sensitivities/fluence thresholds that we compute are referenced to the source's transit through the FWHM of CHIME/FRB's beam at 600 MHz. Bursts that are detected outside of this exposure are not counted in the burst rate calculation. Especially due to the non-Poissonian nature of repeaters, this occasionally leads to significant reduction in the total number of observed bursts considered in the rate calculation, for example all of FRB~20210601A's nine bursts were detected outside of the FWHM at 600 MHz (Mckinven et al. in prep.). Circumpolar sources, which are the sources above a declination of 70 deg, will have additional exposure because of their secondary lower transits, 12 hr after the upper transit, through CHIME/FRB's synthesized beam. Bursts that are detected in the lower transit are not considered in this work unless explicitly mentioned otherwise (e.g., for our repeater fraction analysis) given CHIME/FRB is less sensitive to sources in their lower transit compared to when they are in their upper transit. 
We report the median value from the 95\% highest posterior density interval of burst rates in Table \ref{tab:rn4prop}. This interval is estimated from their observed counts, Poisson uncertainty, and our estimate of the exposure and associated uncertainty on the exposure. The asymmetric uncertainties on the burst rates reported in Table \ref{tab:rn4prop} correspond to the lower limit and upper limit of the same 95\% highest posterior density interval. 
We report the median estimated burst rate, even when the rate estimation is based on the observation of zero bursts. 

\begin{figure*}
    \includegraphics[width=\textwidth]{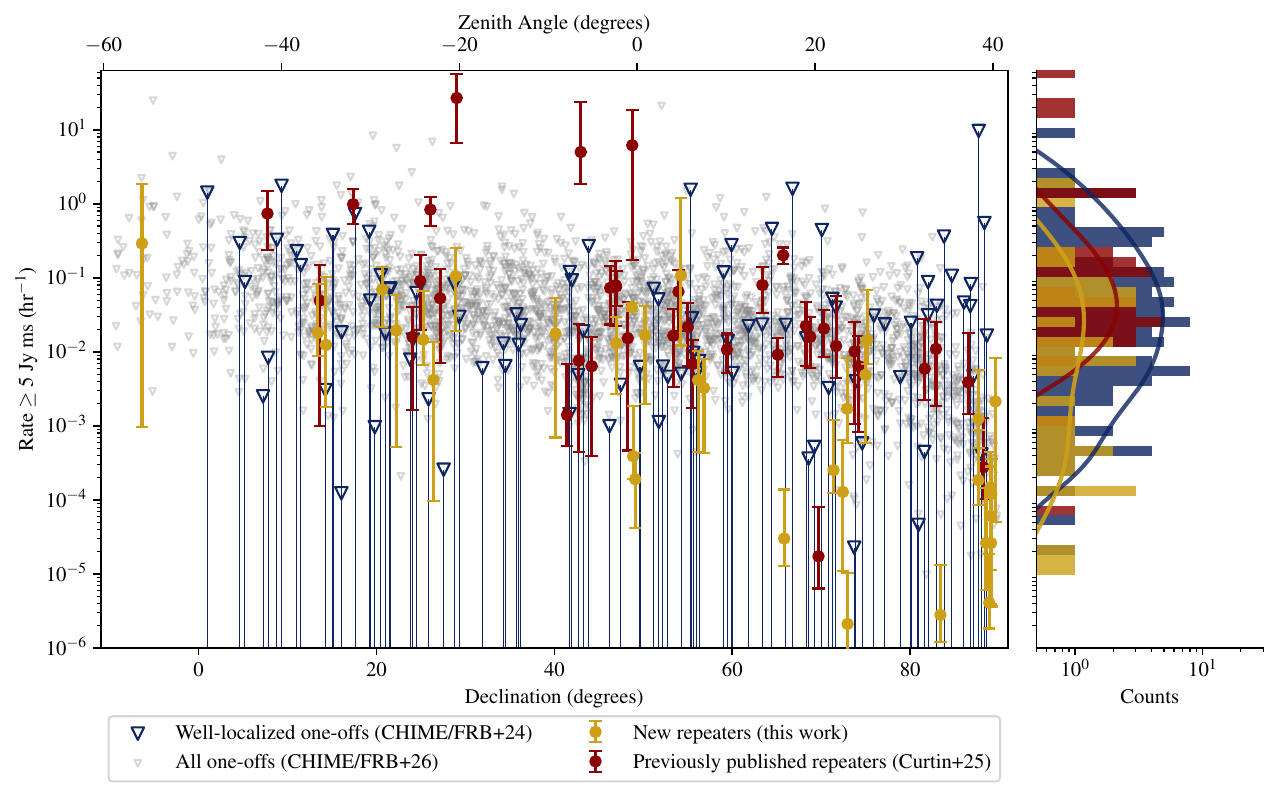}
    \caption{\label{fig:rates} \textit{Left panel:} Burst rate above 5 Jy ms for repeating FRBs from CHIME/FRB vs. source declination. Previously published repeaters \citep{alice,cat2} are plotted in red and repeaters newly reported in this work are plotted as gold dots. In both cases, their error bars correspond to 95\% uncertainty regions assuming Poisson statistics in number of detections and the approximately Gaussian uncertainty in the source exposure.  The 95\% upper limits implied from as-yet non-repeating, or one-offs, from sources reported in \cite{basetcati,cat2} are shown with open blue and gray triangles for the full sample and the arcminute-localized sample respectively. The well-localized CHIME/FRB one-offs from \cite{basetcati} will have smaller upper limit constraints on average, because these estimates are dominated by uncertainty in the exposure, which better localization reduces. For all plotted rate constraints, the total burst rate at a given sensitivity (which varies as a function of declination) has been adjusted to a 5 Jy ms fluence threshold, assuming a $-1.5$ power-law energy index (see \S \ref{sec:rates} for more detail). The three highest rate repeaters are FRBs 20200223B, 20201221B, 20191106C. \textit{Right panel:} Histogram of the 95\% upper limit burst rates above 5 Jy ms for the different populations colored as in the left panel. A Gaussian kernel density estimate (KDE) of the PDF is overlaid for each population; in each case the bandwidth of the KDE was selected using Scott's rule \citep[equal to the number of data points to the power of -1/5 for 1D data; ][]{scott1992multivariate}. The upper limits implied from the observations of as-yet non-repeating FRBs appear largely consistent with the measured burst rates from repeaters.}
\end{figure*}

The burst rates for the sources first published in this work, and updated burst rates from previously published repeaters \citep{alice}, are plotted in Figure \ref{fig:rates}. \cite{alice} do not report estimated uncertainties on the exposure for these sources, so to ensure the rate samples are comparable, we assume a 10\% uncertainty on the exposure and calculate those rates and associated uncertainty as in \S \ref{sec:rates}. This is a reasonable approximation for the average exposure uncertainty for CHIME/FRB sources with localizations from baseband data. We additionally update rates for the Cat2 time period for FRB~20180916B, FRB~20201124A (which are plotted as `previously published repeaters' in Figure \ref{fig:rates}), and FRB~20220912A (which is plotted as a `new repeater' in Figure \ref{fig:rates}, although it was initially announced by \cite{2022ATel15679....1M}, given it was first discovered after the RepCat3 period) in the Appendix Table \ref{tab:extrarates}. 
For comparison, we show the implied 95\% upper limit on rate for apparent non-repeating bursts from Cat2 computed in the same manner. We highlight sub-arcminute-localized CHIME/FRB one-offs from \cite{basetcati} because the upper limits are dominated by uncertainty in the exposure. The uncertainty in the exposure is reduced with increased localization precision.

\subsection{Correlations} \label{sec.correlations}

\begin{figure*}
    \centering
    \includegraphics[width=1\textwidth]{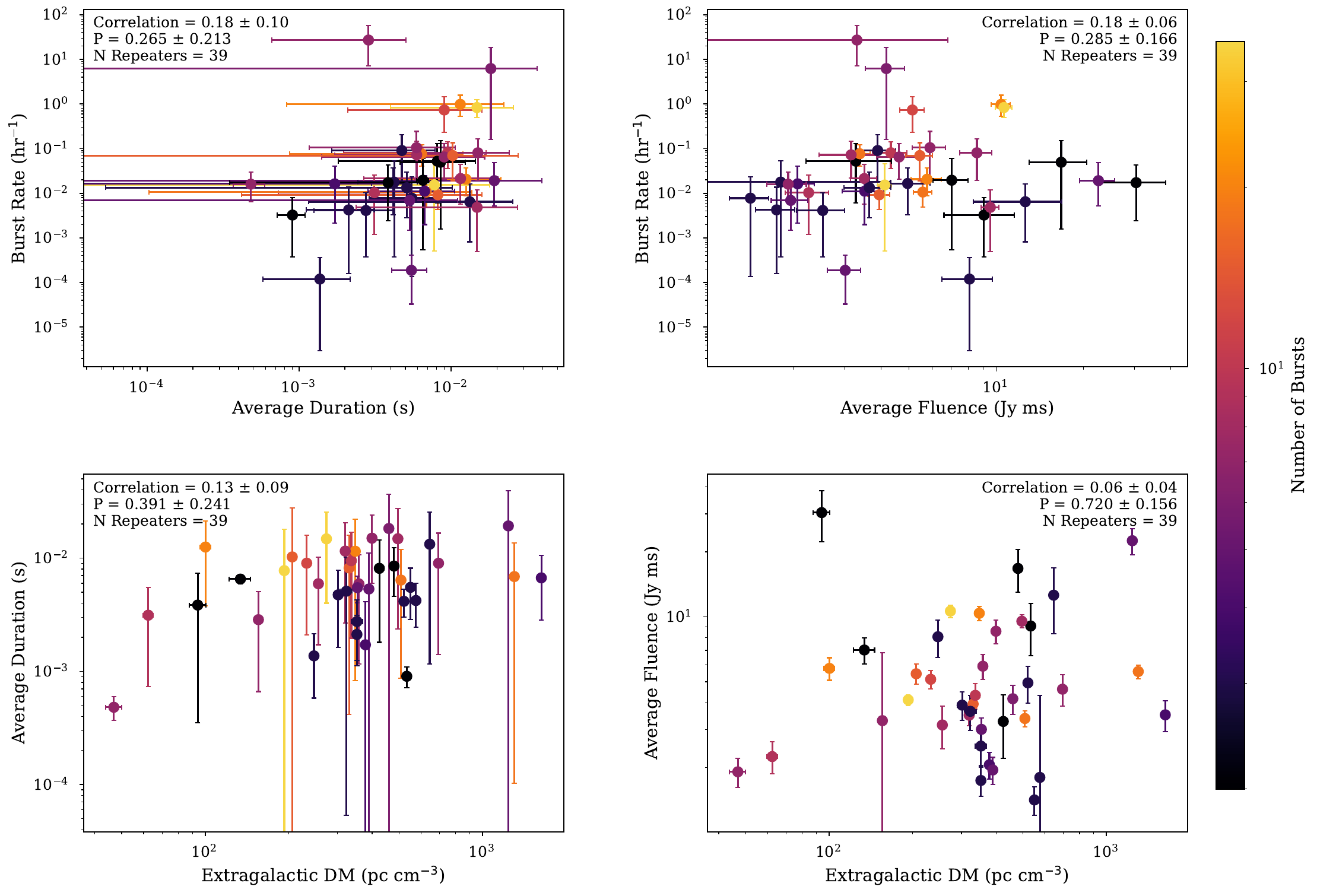}
    \caption{\label{fig:correlations}Correlation plots between burst rate, fluence, burst duration, and the extragalactic DM for the repeating FRBs in our sample. Burst duration and extragalactic DM are given in the observer's frame. Following the approach outlined in \S \ref{sec.correlations}. Only repeaters with at least two bursts detected are included. The quoted Spearman rank correlation coefficients and corresponding p-values represent the median values obtained from 1000 MC realizations, in which the measured quantities are resampled from their uncertainties. The uncertainties in the correlations are quantified using the median absolute deviation from the median (MADFM).}
\end{figure*}

We searched for correlations between source
properties derived from the burst population of each source.
Figure~\ref{fig:correlations} shows combinations of the burst rate,
average burst duration, average burst fluence,
and extragalactic dispersion measure (DM). 
Our methodology broadly
follows the correlation analysis framework described by \cite{RN3} and \cite{alice}. The plotted values of burst duration are in the observer's frame, and the burst rate is based on the upper transit only for  sources with both upper and lower transits (with declinations greater than 70 degrees). 

We restrict our analysis to repeating FRBs with at least two detected
bursts from the sample, that have non-zero burst rates, and further only to bursts that fall within the formal exposure region. After applying this cut, our final sample consists of 39 distinct repeating FRB sources. For bursts with multiple components, we define the total burst duration as the temporal extent of the burst envelope, measured as the arrival time of the last component plus half FWHM, minus the arrival time of the first component minus half of its FWHM. For quantities that
exhibit intrinsic burst-to-burst variation (e.g., duration and
fluence), the plotted error bars represent the standard deviation of a source’s burst property, rather than the uncertainty on measurement.
We use the Cat2 burst DMs from \texttt{fitburst}, which maximize S/N.
We define the extragalactic dispersion measure (DM) as the observed DM after subtracting both the Galactic contribution predicted by the NE2001 electron density model \citep{ne2001} and an additional $30\,\dmunits$ to account for the Milky Way halo \citep{dolag,cookdmhalo}. We assume a $20\%$ uncertainty on the NE2001 Galactic DM contribution,
which is used as the error for the extragalactic DM.

To assess the presence of correlations between repeater properties, we
use the Spearman rank correlation test \citep{Spearman1904}
as implemented in \texttt{scipy.stats.spearmanr} \citep{2020SciPy_virtanen}. A correlation is considered statistically
significant if the Spearman $p$-value satisfies $p < 10^{-3}$. Uncertainties on the correlation coefficients and $p$-values are estimated using an MC approach as performed by \cite{alice}. For each correlation,
we perform 1000 simulations in which repeater-level quantities are drawn from the per-repeater uncertainty regions, assuming Gaussian distributions, aside from the burst rate for which we assume a uniform distribution. For each realization,
we recompute the Spearman correlation coefficient and associated
$p$-value. We report the median of these distributions, along with the
median absolute deviation from the median (MADFM), as our measure of
uncertainty, in Figure~\ref{fig:correlations}.  We report no significant correlations, and our results are broadly consistent with the results of RepCat3 and \citet{alice}.

\subsection{Temporal DM Evolution}

\label{sec:deltadm}
First, for each of the new repeaters from both the gold and silver samples, we searched for additional bursts in Cat2 within one degree of angular separation of their positions (roughly the uncertainty on our least constrained positions) and 40 $\dmunits$ of their mean DMs \citep[roughly the maximum expected variation over the duration of the survey based on measurements of other variable repeaters; ][]{2023MNRAS.526.3652K,2025arXiv250715790W,2025arXiv251011352S,2025arXiv250916374O,NIU202676,pandhi26,thomasr117}. This search is to recover any additional bursts from the sources which may have been missed by the clustering algorithm due to significant DM evolution over time. We found no compelling additional candidates with these searches. For those repeaters with four or more bursts\footnote{Some FRBs and pulsars have demonstrated variability in their DMs beyond that attributable to measurement error, due to e.g., profile variations or ISM variations. As such, we choose to only perform this test for repeaters with four or more observed bursts to reduce our chances of over-interpreting a snapshot of stochastic behavior as deterministic \cite[See, for example, ][]{2013MNRAS.435.1610P,2017ApJ...841..125J,2018MNRAS.479.4216M,2020A&A...644A.153D,2024MNRAS.530.1581K,alice}.}, both from our gold sample and all previously published CHIME/FRB repeaters (40 sources total), we then
search for evidence of deterministic DM variation by fitting both a constant model (a weighted mean) and a weighted linear model via maximum likelihood estimation to the measured DMs as a function of time. We evaluate the goodness of fit of both best-fit models using the Bayesian Information Criterion (BIC), which penalizes the additional complexity of the linear model compared to a constant model. When the difference between the BIC for the best-fit constant model and the best-fit linear model is greater than 10, it is considered strong evidence that the linear model is preferred. 

We find 26 of the sources with four or more bursts have a $\Delta$BIC $> 10$. In order to further diagnose the significance of these models, we simulate DMs from a normal distribution with the same mean and variance as the sample of DM measurements for a given repeater. We keep the detection arrival times and DM uncertainties fixed, and record how frequently we find the best-fit linear model of the simulated data has an equal, or more significant, BIC. This simulation helps distinguish between cases where the linear models are preferred due to stochastic variation which aligns in an approximately linear fashion by chance. For eight sources we find a tentative linear DM variation, where we find a $p$-value below 1\% using the simulation (set such that the $p$-value corresponds to a family wise error rate of 0.5 if one applies the Bonferroni correction using our 40 searched sources as a trials factor).

\begin{figure*}
    \centering
    \includegraphics[width=\linewidth]{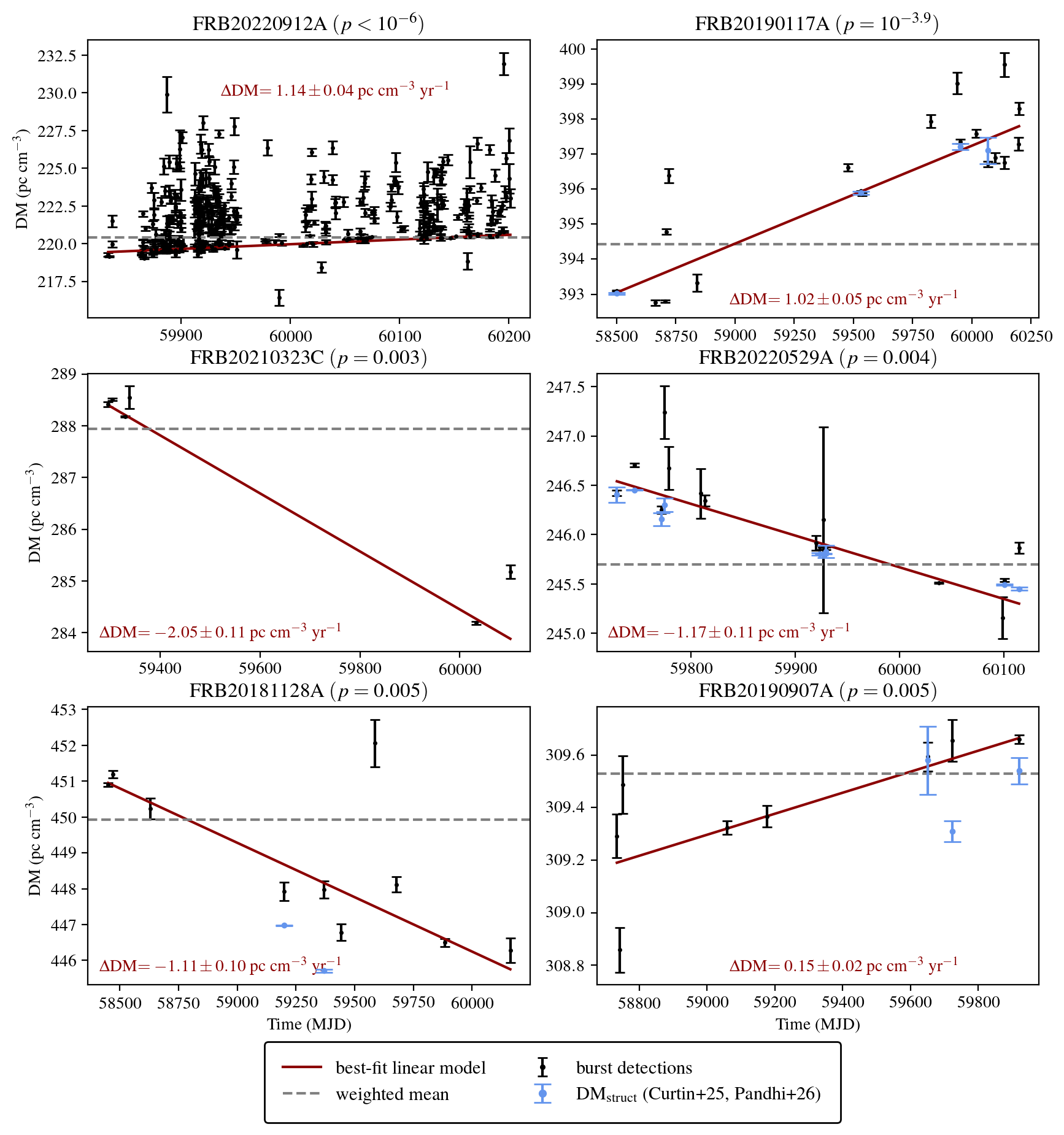}
    \caption{\label{fig:deltadm} DM versus burst time of arrival for sources (black data points) for which the best-fit linear model (red solid line) is strongly preferred to the best-fit constant model (dashed grey line) and corresponding simulation-based $p$-value is smaller than 0.001 (which is equal to a 50\% family-wise error rate across the sample given our trials factor of 40 searched repeaters) and DM variation is not dominated by a single burst. Sources are ordered by increasing $p$-value or decreasing significance. FRB~20220912A and FRB~20220529A have been explored using higher time resolution data by \cite{thomasr117} and \cite{pandhi26}, respectively, and in each case, the variation was confirmed to be significant. Each panel is annotated with the best-fit slope in red text. Where previous measurements of the structure maximizing DMs based on higher time-resolution raw-voltage data for these sources have been published, we superimpose those measurements in blue \citep{alice, pandhi26}.}
    \label{fig:tentativedm}
\end{figure*}

For two of these sources, FRB~20210209C and FRB~20220505A, the variation is dominated by the inclusion of a single burst DM measurement outlier, and upon inspection of their dynamic spectra, it is clear in both cases that the burst morphologies do not have sharp enough features to distinguish between the weighted mean DM of the source (see \S\ref{sec:deltadmdisc} for more discussion on repeater burst morphology and DM bias). We present the DM versus time plots of the remaining six sources, their best-fit linear models, and their associated $p$-values in Figure \ref{fig:deltadm}. We also include the statistical uncertainty on the slope in Figure \ref{fig:deltadm}, estimated from the covariance matrix scaled by the reduced $\chi^2$. 

These six sources include FRB~20220529A and FRB~20220912A, whose linear monotonic DM variation have been explored previously using higher-time-resolution data \citep{thomasr117,pandhi26}. 
The remaining four sources are newly discovered as having tentative monotonic, linearly varying DMs with time and include both positive and negative slopes, corresponding to increasing and decreasing DMs with time respectively. The observed linear variation has absolute magnitudes spanning $0.15-2.1$ pc cm$^{-3}$ per year, and demonstrate this variation spanning one to nearly three years. 

We emphasize that the lower time resolution total intensity data used here to estimate DM are much more prone to degeneracies between DM and morphological burst features and this effect is not entirely characterized by our variation simulations above. This is discussed in more detail in \S\ref{sec:deltadmdisc}. We super-impose published structure-maximized DM measurements from the higher-time-resolution CHIME/FRB raw-voltage data where available \citep{alice,pandhi26}. Generally these more accurate DMs strongly support (in the case of FRB~20190117A) or at least do not rule out the proposed variation.

Only one of the candidates is significant above the 3$\sigma$ level ($p$-value $< 0.003/40 =  7.5\times 10^{-5})$ using this simulation, and that is FRB~20220912A. The potential variation of FRB~20220912A was first described by \cite{2024MNRAS.534.3331K} and detected at the $2.3 \sigma$ level by \cite{thomasr117}. \cite{thomasr117} explore this variability using CHIME/Pulsar data, which has higher time resolution and sensitivity than those data we are considering here.
FRB~20220529A, that was shown by \cite{pandhi26} to have a significant linear DM variation using this same test but with a longer spanning and higher-resolution baseband dataset, is recovered here as an ambiguously significant linearly varying source from this test. This is promising for the remaining four sources with a tentative monotonic linear DM variation, and we will revisit the measurement of these sources' DMs with time with the increased precision offered by the raw voltage data and with any additional detections outside of the Cat2 period in an upcoming analysis.

Using the search detailed above, we additionally checked the position of known repeater FRB~20121102A for additional bursts \citep{R1}. CHIME/FRB has only detected FRB~20121102A once \citep{2019ApJ...882L..18J} but is known to have a significantly evolving DM (changing $12~\dmunits$ in $\sim$10 years; \citealt{2019ApJ...876L..23H,2025arXiv250715790W,2025arXiv251011352S}). We find no additional candidates for repeat bursts from this source. 

\subsection{Comparison with previously identified repeaters}

\begin{figure*}
    \includegraphics[width=\textwidth]{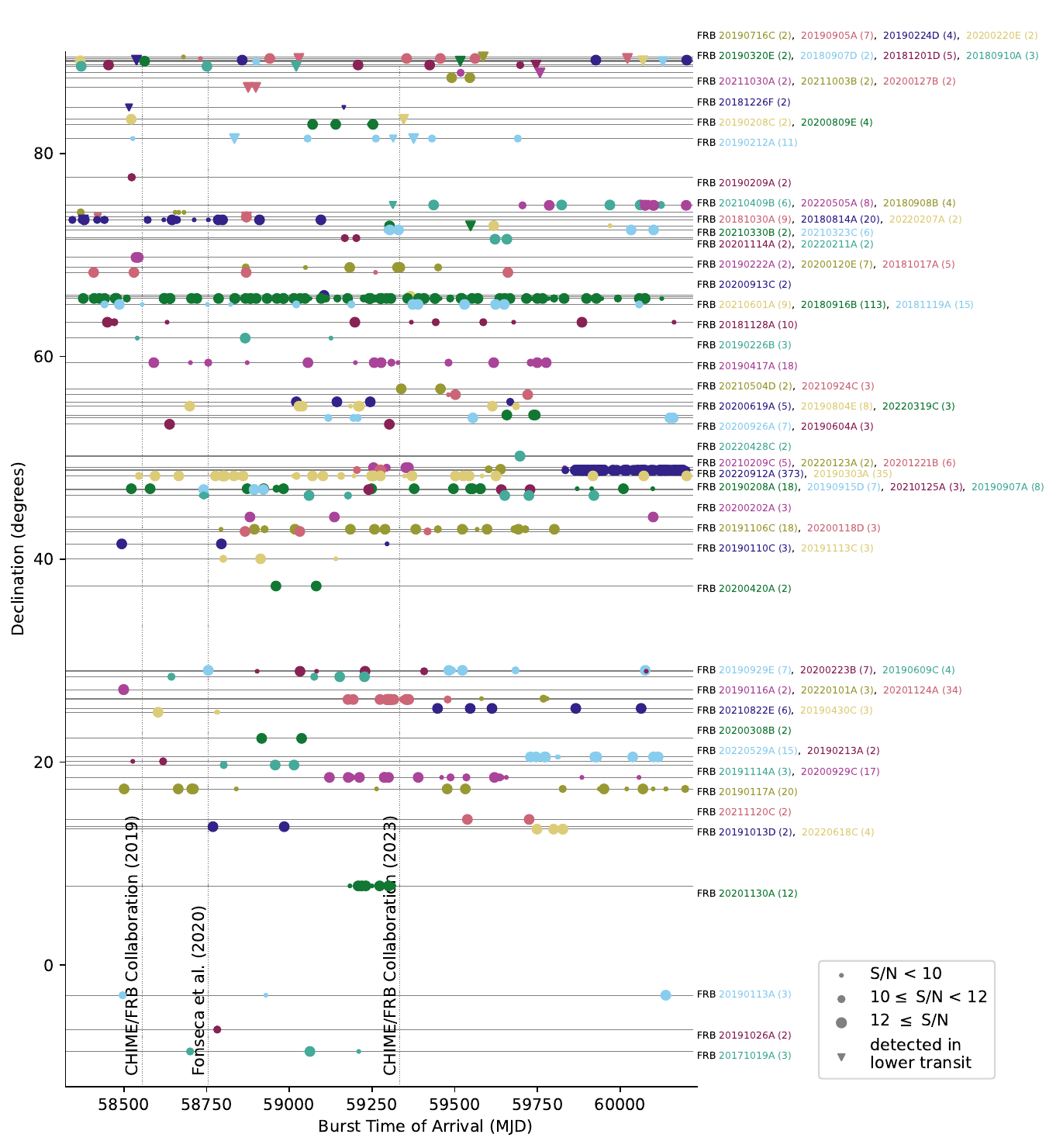}
    \caption{\label{fig:weatherreport} Burst detection time versus declination of all of CHIME/FRB's discovered repeaters. Marker size indicates the S/N of the burst detections. Strong evidence of clustered burst detections, or evidence of enhanced activity, is apparent for many sources, for example FRB~20220912A, FRB~20201124A, and FRB~20201130A (near Decl. $\sim$48, 26, and 8 degrees, respectively). Individual sources are identified with a gray horizontal line at their declination, and listed on the right with their number of observed bursts included parenthetically beside their names. Figure updated from \cite{RN3}.}
\end{figure*}

The estimated burst rates of the gold sample sources are smaller, on average, compared to those found by \cite{alice} for previously published repeaters, using the same Cat2 timeframe, as can be seen in Figure \ref{fig:rates}. 
Under the hypothesis of stationary Poisson rates, it is expected that CHIME/FRB would discover repeaters with increasingly smaller burst rates as time went on. However, given the highly clustered behavior and apparent finite lifetimes of repeaters, the significant overlap we see in the populations is unsurprising. For example, we do not detect any additional bursts from 41\% of previously known repeaters in the increased exposure between Cat2 and RepCat3 (RepCat3 included source observations only up to 2021 May 1). We continue to observe strong evidence of fluctuating activity levels for both the new and previously published repeaters, best illustrated by their burst detection timelines, plotted in Figure \ref{fig:weatherreport}. We also have a larger fraction of very high declination repeaters in our gold sample compared to the previously published repeaters. This is likely due to increased sensitivity offered by the new methodology for estimating \pcc{} at high declinations. Having more sensitivity at high declinations will also lead to the detection of repeaters with lower inferred burst rates, given they are circumpolar or near circumpolar and hence have the highest exposures.

The extragalactic DM distributions of previously published repeaters and the gold sample of this work are consistent with being drawn from the same underlying population: An Anderson-Darling test of the two samples finds that they are consistent with coming from the same underlying population with a $p$-value of 0.74.

\begin{figure}
        \centering
        \includegraphics[width=1\linewidth]{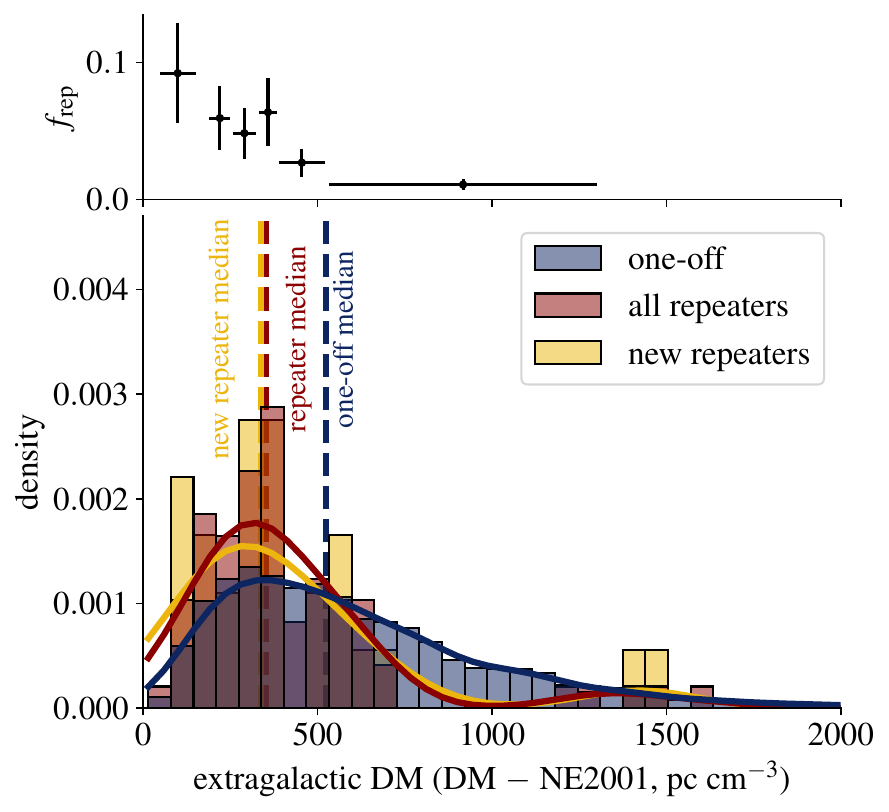}
        \caption{\textit{Bottom panel:} Extragalactic DM histograms for the apparent one-off population (dark blue histogram), the repeaters newly reported in this work (gold histogram), and all previously published repeaters from CHIME/FRB (red histogram). Extragalactic DMs are reported in the observer's frame. The medians of each sample are shown in dotted vertical lines of their respective colors, to demonstrate consistent distributions between the previous and new repeaters, but discrepancies with the one-off populations. An AD test on the repeaters and one-offs find the distributions inconsistent with being drawn from the same underlying distribution, with estimated $p$-value dmaller than $1\times10^{-5}$. \textit{Top panel:} $f_{\text{rep}}$, the fraction of FRB sources observed to repeat, in six extragalactic DM bins defined such that there are an equal number of repeating FRB sources in each. This fraction is necessarily a lower limit on the true fraction of repeaters, given that our measured repeater rates would be largely consistent with only detecting one FRB from a source, and given CHIME/FRB is biased against repeater-like burst morphology.}
        \label{fig:dms}
\end{figure}

\section{comparisons to apparent one-off 
population}
\label{sec:r_v_o}
\subsection{Inclusion Criteria}
In the following analyses, we enforce a few selection criteria to ensure that the exposure of FRBs is well defined and corresponds to a meaningful sensitivity, and so that we include only bursts above an S/N threshold above which CHIME/FRB is reasonably complete. Specifically, we only consider bursts with S/N $\geq 12$ and that were 
detected within the FWHM of CHIME/FRB's formed beams at 600 MHz, which is where the exposure is defined. At S/N below 10, we have different callback criteria for repeat bursts compared to  sources discovered for the first time, making the samples incomparable (\citetalias{cat2} \citeyear{cat2}). Similarly, events below S/N of 12 are more likely to be classified as noise upon
human inspection. We also only consider sky exposure and bursts detected from the upper transit for declinations above +70 deg.  

\subsection{Repeater fraction}
\begin{figure*}
    \includegraphics[width=\textwidth]{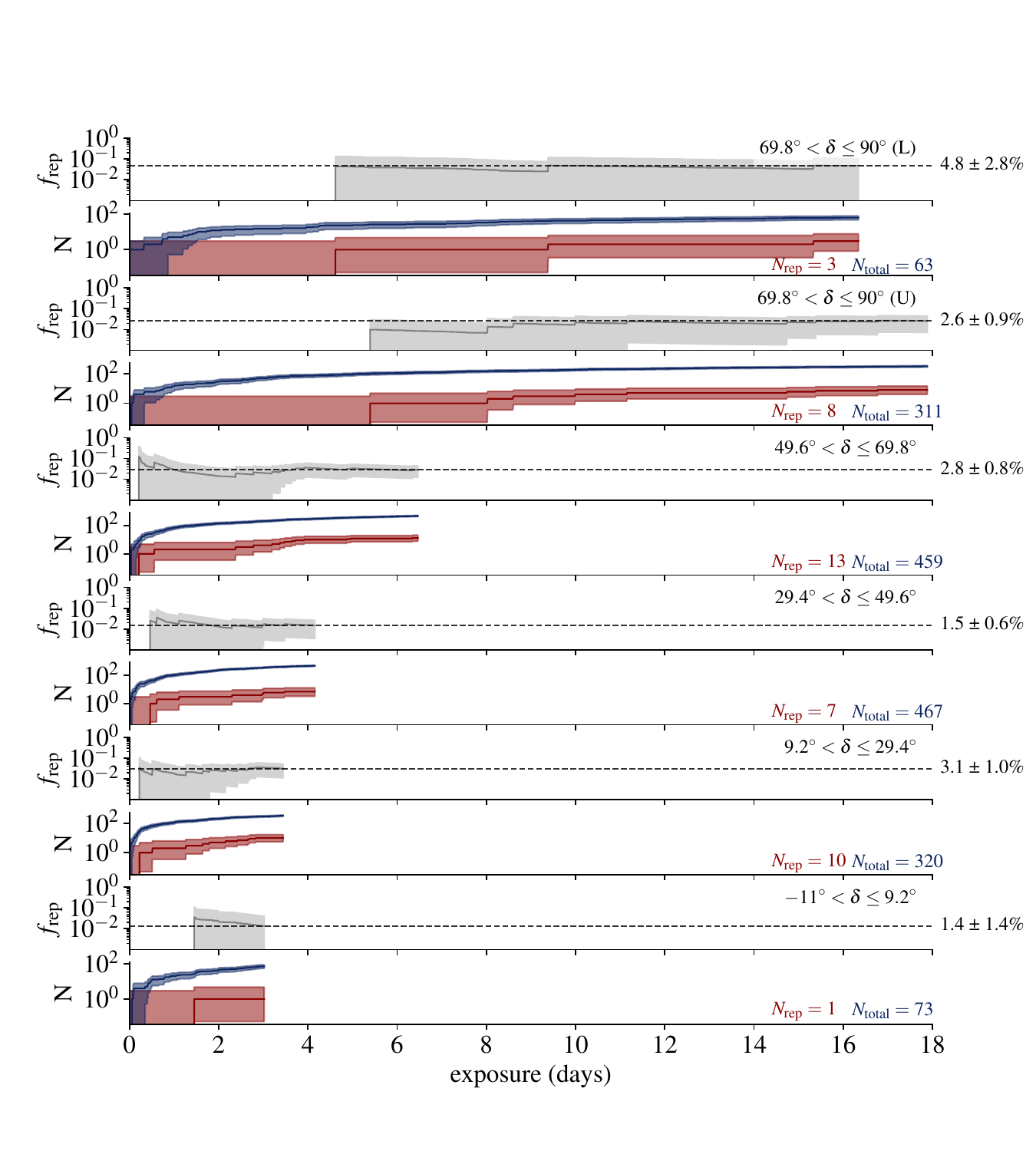}
    \caption{\label{fig:fraction} Repeating FRB fraction (number of FRB sources that have been observed to repeat divided by the total number of observed sources) as a function of average total exposure in different declination bins (grey line). Declination bins are indicated in the top right corner, with U/L separating the upper and lower transit of the top bin, respectively. In each pair of panels, the first represents the fraction of FRBs that have repeated, and the second shows the total number of repeaters (red line) and total number of sources (blue line) detected in each bin as a function of exposure in that declination bin. In all cases, the lighter shaded region around the mean estimates of these values represents the 95\% confidence interval, and the reported errors correspond to the 95\% uncertainty ($2 \sigma$ Gaussian equivalent). These fractions are necessarily lower limits of the true repeating fraction of FRBs.}
\end{figure*}

 With the selection criteria above applied, we compute the observed repeater fraction as a function of time. Following RepCat3, we consider the time of `discovery' of a repeater as the exposure value when we detect a second burst (that satisfies all the population inclusion criteria) from that source. The number of total FRB sources, $N_{\text{FRB}}$, and the number of observed repeaters, $N_{\text{rep}}$, are assigned Poisson counting errors $\sigma_{N_{\text{FRB}}}$ and $\sigma_{N_{\text{rep}}}$, respectively. The repeater fraction 

\begin{align}
    f_{\text{rep}} & = \frac{N_{\text{rep}}}{N_{\text{FRB}}}
\end{align}
has approximate uncertainty, adopted from RepCat3,

\begin{align}
    \sigma_{ f_{\text{rep}} } & = \sqrt{\left(\frac{\sigma_{N_{\text{rep}}}}{N_{\text{FRB}}}\right)^2 + \left( \frac{N_{\text{rep}} \sigma_{\text{FRB}}}{N^2_{\text{FRB}}} \right)^2}.
\end{align}
 We compute this for six independent declination bins because declination is the biggest predictor of our sensitivity. One could imagine that surveys with different sensitivities and exposures would probe different regions of a populations luminosity function, and hence would have the potential to obscure an otherwise-observable plateau or significant change.
 We show these repeat fraction curves in Figure \ref{fig:fraction}. The observed repeater fraction seems to tend to an equilibrium value 1--5\% in all bins, all consistent with one another when considering the 95\% confidence range. In particular, our observed repeater fraction does not show evidence of asymptotic growth towards unity or a decrease with time. The overall observed repeater fraction is measured to be $2.4\pm0.4\%$. 

The observed repeat fraction is necessarily a lower limit on the true fraction of FRB sources which truly repeat for two reasons. First, there is no evidence that we have begun to probe the lowest possible repetition rates of repeaters, so it is plausible a large fraction of the sources are repeaters but are simply not active enough to expect multiple detections in our current exposure. Second, CHIME/FRB is biased against detecting the narrower bandwidths and wider durations of repeater bursts (compared to one-offs), given such repeaters will, on average, have lower S/N than one-offs for bursts which represent the same total fluence. 

The observed evolution of $f_{\text{rep}}$ with time can be informative for the underlying population. If all FRBs repeat on timescales much shorter than their active lifetimes, one might expect that the observed repeating fraction would asymptotically tend to unity with enough exposure. Similarly, if repeaters are only a small fraction of the total observed FRB population but still repeat on timescales much shorter than their active lifetimes, one might expect that with enough exposure you would detect the majority of the observable repeaters and the repeating fraction could begin to decrease. \cite{frbpoppy} explored the behavior of the observed repeater fraction versus time while varying the fraction of sources that repeat in the population, as well as distribution of repetition rates. In an idealized observation scenario, \cite{frbpoppy} find these observed repeater fraction curves look very different in a population where 100\% of sources repeat with a fixed rate of 0.1 day$^{-1}$ versus 50\% repeat with the same repetition rate and 50\% do not repeat at all. However, \cite{frbpoppy} also find that a CHIME-like observing beam pattern makes these two scenarios appear very similar, because one cannot distinguish a turnover in $f_{\text{rep}}$ over time, at least on the hundreds of days timescales shown by \cite{frbpoppy}. 
By forward modeling the FRB population and correcting for the declination-dependent exposure times of CHIME/FRB,  \cite{2024MNRAS.52711158Y} find that the source count evolution from Cat1 is consistent with arising from a true/underlying repeater fraction that exceeds 50\% with 99\% confidence. The observed repeater fraction from Cat1 and RepCat3 are similar to that seen in the Cat2 sample (this work), and qualitatively there are not significant differences in the observed repeater fraction evolution compared to that presented in RepCat3, and so we expect such forward modeling would conclude a similarly high true repeater fraction. A full reanalysis is outside of the scope of this work, but would provide a more quantitative statement.

\subsection{Rates}
\label{sec:rates_r_v_o}

Qualitatively, in Figure \ref{fig:rates}, it seems as though the distribution of repeater rates is consistent with the distribution of upper limits implied from observations of as-yet one-offs. The least active repeaters in our sample have estimated rates below the lowest upper limits we place on apparent one-offs and, while the average repeater rate ($10^{-0.22}$ hr$^{-1}$ above 5 Jy ms) is higher than the average upper limit placed on the well-localized apparent one-off sample ($10^{-0.57}$ hr$^{-1}$ above 5 Jy ms), they share much of the same phase space, suggesting the observations are still consistent with a single population of repeating FRBs with a distribution of source repeat rates spanning many orders of magnitude. This is in contrast to a clear bimodal distribution of rates and upper-limits, that would be more suggestive of a second population of truly non-repeating FRBs. 
It is still possible that such a clear bimodality in the populations would begin to emerge with orders of magnitude more survey exposure.  There are a few sources which are visual outliers that have repetition rates an order of magnitude or two higher than the upper limits on repetition rates of the apparent one-offs, but the majority of measured repeater repetition rates are not distinct from the non-repeater upper limits. Interestingly, the three highest source rates correspond to FRBs 20200223B, 20201221B, and 20191106C, none of which are particular outliers in total number of burst detections, demonstrating how much the sensitivity and exposure vary from sight line to sight line and their affect on the rate estimates. FRB~20220912A, for example, is a distinct outlier in total burst detections (210 bursts in Cat2) and has been classified as hyperactive in other works (e.g. \citealt{2024MNRAS.534.3331K,2024MNRAS.529.1814H}) but does not stand out from other repeater rates for CHIME/FRB when corrected for sensitivity and exposure and averaged over the duration of Cat2. We caution, however, that we are assuming a universal power-law cumulative burst energy distribution with index of $-1.5$ to scale these rates, and this is not appropriate for every source.

It is difficult to statistically compare the population of upper limits to the measured rates, given the former are left-censored versions of the observable we would like to compare. However, the qualitative agreement of the two populations is not surprising, given that CHIME/FRB still detects second bursts for repeaters that originally were detected in the first years of our experiment, suggesting we have not, at our current exposure, probed the least active repeaters. Given we are biased against detecting sources with repeater-like morphology, the repeater rates we report are biased low. Future modeling that accounts for this detection bias, and/or orders of magnitude more exposure, could begin to reveal any bimodality in these distributions. 

\subsection{Population Modeling}

As was first reported in RepCat2 and RepCat3 as a tentative detection, there is a discrepancy between the distributions of DMs for apparent one-off and repeating FRBs. The apparent one-offs have a median and mean extragalactic DM of 524.7 $\dmunits{}$ and $629.4$ $\dmunits{}$, respectively. The observed repeaters have a median and mean extragalactic DM of 353.0 $\dmunits{}$ and $424.1$ $\dmunits{}$, respectively. The updated DM distributions of the two samples can be seen in Figure \ref{fig:dms}. The discrepancy has gone from marginal to decisive, with an Anderson-Darling $p$-value smaller than $10^{-5}$ for being drawn from the same underlying population.  We also show the observed repeating fraction as a function of extragalactic DM in Figure \ref{fig:dms}.

A question naturally arises: does this discrepancy imply that the two samples are being drawn from different distributions, and hence can we assume at least some apparently non-repeating FRBs come from a separate population to known repeating FRBs? Instead of intrinsically distinct populations, it could be at least partially be explained as an observational bias,  given that in order for a source to be deemed a repeater in this sample, we must detect at least two bursts from the source. This criteria naturally lends itself to a bias towards lower DMs: FRBs with low DM must also have (relatively) low $\dmcosmic$ and thus tend to be closer on average. Assuming that all FRBs have roughly the same underlying luminosity distribution, closer sources have a much greater chance of emitting two or more detectable bursts \citep{frbpoppy}. 

There are additional observational effects that bias CHIME/FRB away from detecting repeaters. Repeaters are, on average, narrower banded and broader in temporal duration \citep{pleunismorph}. This phenomenon has the effect that for a given energy/fluence, an average repeater burst will have lower peak signal-to-noise than an average one-off burst, and hence be more difficult to detect. This again leads to a bias towards only detecting the brighter sources as repeaters, and hence, using the same argument above, a bias towards repeaters being detected at lower DMs.

\cite{2023PASA...40...57J} (henceforth J23) explores the discrepant DM distributions, in addition to other observations, of Cat1 and RepCat1\&2, in the context of population modeling using the open source \texttt{zDM} package developed by \cite{zDM}\footnote{\href{https://github.com/FRBs/zdm}{https://github.com/FRBs/zdm}}, which accounts for instrumental selection effects\footnote{In so far as our instrumental selection effects are characterized in Cat1 and the associated synthetic injection campaign \citep{2023AJ....165..152M}, which mainly accounts for declination dependent sensitivity and exposure, as well as our DM selection, but not our selection against repeater-like morphology.} and cosmology to calculate the number of FRBs detected by a given survey for some underlying FRB energy function and redshift distribution. 
J23 find that, assuming that $50-100$\% of apparent one-off FRBs are attributable to repeaters with a power-law distribution of FRB repetition rates, one can recover a distribution of source declinations, DMs, observed repeaters, and numbers of repeats for each repeater which describe our Cat1 sample well. 

\begin{figure}
    \centering
    \includegraphics[width=\linewidth]{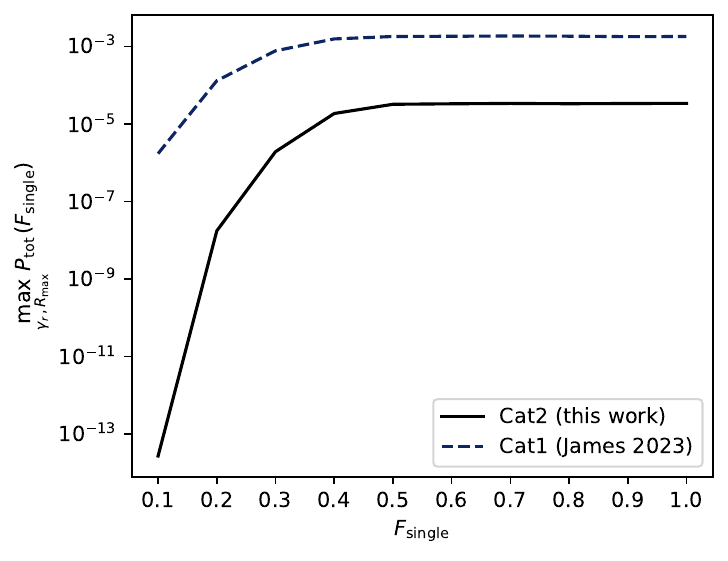}
\caption{\label{fig:clancyprob} The maximum probability of the J23 repeating population model as a function of the fraction of as-yet one-off FRBs observed by CHIME/FRB that are assumed to be attributable to repeaters, $F_{\text{single}}$. The dotted line shows the original values found when applied to the Cat1 sample by J23. The solid line shows the results for the model fit to Cat2 data. In both cases, fractions between 0.5--1.0 have higher probabilities than lower fractions, suggesting the DM, Declination, and rate distributions of repeaters and one-offs are equally well explained when 50--100\% of our as-yet one-offs are attributable to repeaters, assuming the J23 population model.}
\end{figure}

To check what range of underlying repeater fractions is still consistent with the Cat2 sample and its repeaters, we rerun the simulations from J23. The few changes that we make to the original analysis are (\textit{i}) reducing the assumed value of DM$_{\text{halo}}$ from 50 to 30 $\dmunits$ \citep{dolag,cookdmhalo} to ensure that all FRBs have positive estimated extragalactic dispersion measures and (\textit{ii}) enforcing the S/N criterion of our sample (only including bursts with S/N $> 12$, although we confirmed the dataset with S/N $> 12$ produced consistent broad conclusions as when the entire sample was used, as with either DM$_{\text{halo}}$ value). 
In J23, the repeater population is described by three model parameters: the maximum repetition rate  $R_{\text{max}}$, the minimum repetition rate $R_{\text{min}}$, and the power-law index, $\gamma$, that connects them to describe the total distribution of repeat rates in the population. In the J23 formalism, $\gamma$ and $R_{\text{max}}$ are varied and $R_{\text{min}}$ is optimized for a given $\gamma$ and $R_{\text{max}}$ pair to reproduce the number of observed CHIME/FRB repeaters and some fraction, $F_{\text{single}}$, of the total singles burst rate. $F_{\text{single}}$ is the fraction of as-yet non-repeating FRBs attributable to repeaters, and different values are tested.  Among all tested values of $\gamma$ and $R_{\text{max}}$, the probability of the best-fit parameters is shown as a function of assumed $F_{\text{single}}$ in Figure \ref{fig:clancyprob}. The probability which is maximized to find the best fitting $R_{\text{max}}$ and $\gamma$ for each $F_{\text{single}}$ is a joint probability, $p_{\text{tot}}$, which is a product of the Poisson probability of observing our 80 repeaters, the likelihood of the observed histogram of the number of repeats per repeater as estimated from MC realizations from the model, the agreement of the DM distributions predicted by the best fit model compared to that observed in Cat2, and the Declination distribution agreement between the model and Cat2. The agreement of the model is evaluated via the $p$-value of a KS test for the declination and DM.  
 
In Figure \ref{fig:clancyprob}, we additionally plot the best-fit probability as a function of $F_{\text{single}}$ that J23 found using Cat1 to compare. The peak values of $P_{\text{tot}}$ are smaller in the Cat2 fit than the Cat1 fit, but since both curves are normalized such that $\iint P_{\text{tot}}(F_{\text{single}}=1.0)  \dd \gamma_r \dd R_{\text{min}}=1 $ for their respective fits of $F_{\text{single}}=1.0$, this is primarily  a statement about how strongly different values of $\gamma_R, R_{\min}$ are favored compared to other values tested for each dataset. That value being lower is not surprising given, e.g., the Poisson distribution is broader for higher expected rates. Instead we examine the relative maximum probability value as $F_\text{single}$ is varied.  In both cases, $F_{\text{single}} \in [0.5, 1.0]$ are essentially equally likely, suggesting the relative number of repeating and as-yet one-off FRBs, and their DM, declination, and burst rate distributions, can be well explained by 50--100\% of as-yet one-off FRBs being attributed to repeaters. While we do not see any change in the $F_{\text{single}}$ which are well-supported, the normalized values at low $F$ drop away from their peak values at 0.5--1.0 more dramatically than found in J23.

\section{Discussion}
\label{sec:disc}
We have reported 30 new repeating FRB sources from CHIME/FRB Cat2.  In addition, from our sample of 80 repeaters detected with CHIME/FRB, we recover rate distributions, observed repeating fraction with time, and population modeling results consistent with a single population of FRBs drawn from a continuum distribution of rates. Within this population, we see no significant, physically meaningful, correlations between source observables. We see significant variation in DM for some of these sources, including six sources for which there is evidence of monotonic, linear evolution. We explore the implication of these results and place them in more context below. 
\subsection{Do All FRBs Repeat?}
 
Deep, targeted follow-up observations of apparent non-repeating sources have only resulted in a detection a handful of times \citep[e.g., the first repeater; ][]{R1}. For some of the cases where these observations fail to detect any subsequent bursts from an apparent one-off, the upper limits on burst rate implied by these observations have been lower than the rates that we have observed for repeaters with CHIME/FRB, assuming stationary Poissonian burst rates that are not dependent on observing frequency and a power-law fluence distribution with index of $-1.5$ \citep{2018Natur.562..386S,2023ApJ...944...70G,2025MNRAS.540.3709U}.
 However, we know all of these assumptions to be false in at least some cases, and the continued discovery of new repeaters with waiting times comparable to the entire duration of the CHIME/FRB survey suggests that we have not yet encountered the lower limit of possible averaged repeat rates from these sources. 

In the gold sample of repeaters from this work, the greatest amount of observed exposure between two subsequent detections of bursts from a repeater is 160.7 days, and the largest elapsed waiting time between subsequent detections is 2.92 years. These are both upper limits on the true waiting time given CHIME/FRB does not observe each position continuously. CHIME/FRB has only detected the first known repeater, FRB~20121102A, once, in the first year of our survey. \cite{2024ApJ...971L..51B} and more recently \cite{2025arXiv250715790W} suggest that this detection rate is expected given the luminosity distribution of this source and given CHIME/FRB's comparatively low sensitivity of 7 Jy ms towards the source, but it serves as a stark reminder that even the non-detection of repeat bursts over our 5+ year survey does not necessarily give us a comprehensive view of a source. Hence, it will likely be difficult to identify a population of truly non-repeating FRBs or conclude that all FRBs eventually repeat through targeted follow-up observations alone.

There are differences between observed repeaters and apparent one offs. First, the phenomenological dichotomy: the average repeater burst is significantly wider in duration and shorter in bandwidth than an average apparent one-off burst. Compact persistent radio sources (PRS) have been observed for a number of repeating sources \citep[e.g.,][]{2017ApJ...834L...8M,2022Natur.606..873N,2025A&A...695L..12B,2026ApJ...996L..16M}, but have not yet been observed for an apparent one-off. Searches for PRS counterparts to apparent one-off FRBs are limited but have been attempted (e.g., \citealt{2025arXiv251202693L}). However, the compactness of these sources have yet to be confirmed and therefore no non-repeating FRB has been associated with a PRS to-date. The hosts of repeating FRBs seem to extend to lower stellar masses than those of the apparent one-off FRBs \citep{2023ApJ...954...80G}. 
Another argument for a separate population of truly one-off FRBs has included energetics, such as in the case of FRB~20250316A (\citetalias{rbfloat} \citeyear{rbfloat}). While still consistent with CHIME/FRB's repeating population in overall rate, observations of FRB~20250316A seem to be in tension with the burst energy distributions of the repeaters where this has been well measured (generally, `hyperactive' repeaters). 

Increasingly it has been observed that the highest energy bursts from repeaters are more common than suggested by the steep power-law of their lower-energy burst distributions \citep{2024MNRAS.534.3331K,2024NatAs...8..337K,2024arXiv241017024O,2025arXiv250916374O}. The luminosity distributions of the highest energy bursts appear to more closely resemble the apparent one-off population, for which population modeling infers a flatter energy distribution with $\gamma \in (-1, 0)$ \citep{2022MNRAS.510L..18J,2022MNRAS.510.1867L,2023ApJ...944..105S} and the maximum energies of apparent one-offs and repeaters appear to be similar ($\sim 10^{42}$ erg; \citealt{2026arXiv260219335S}). However, a broken power-law burst energy distribution is not distance invariant, and thus the description of individual sources using this function would require some knowledge or estimation of their distance (to place the energy where the distribution transitions from the first power-law index to the second).

Measured distances through host associations have historically been limited to a very small population of FRBs. This will change in the new era of routine (sub-) arcsecond localizations from surveys like CHIME/FRB Outriggers \citep{outriggeroverview},  DSA-110 \citep[e.g., ][]{2024ApJ...967...29L}, MeerKAT \citep[][]{2023MNRAS.524.4275J}, and eventually the Canadian Hydrogen Observatory and Radio-transient Detector \citep[CHORD; ][]{2019clrp.2020...28V}, DSA \citep{2019BAAS...51g.255H}, and the Square Kilometer Array \citep[SKA; ][]{2015aska.confE..55M}.  These localizations will also allow consistency checks between the populations via their galaxy types and source positions within them for the respective populations \citep[e.g. ][]{2021ApJ...917...75M,2024Natur.635...61S,2025ApJ...993..119G} and should offer additional constraining power to inferences made from population modeling. 

Given that we can still, under a single population of entirely repeating FRBs, reconcile the fraction of, and the inconsistent DM distributions between, observed repeaters and apparent one-offs assuming the J23 population model, it is clear that the DMs, declinations, relative population size, and number of repeat bursts do not provide evidence for bimodal populations of repeaters and one offs in the current sample. This could be because all FRBs repeat, or simply because we do not yet have the statistical power to differentiate between a bimodal scenario or a continuum population.

However, the J23 analysis, both its original application to the Cat1 and in this work, does not account for the way telescopes are biased against detecting signals with repeater-like spectral and temporal morphology, nor the increasingly ubiquitous observed broken power-law burst energy distributions of repeaters. Both of these effects favor the detection of one-offs compared to repeaters, and thus accounting for them should increase the observed repeating fraction, the average rates of repeaters, and the average DM of repeaters. 

Thus, in order to robustly claim that a single population of repeating FRBs like that modeled in J23 is inconsistent with our observations, the bias due to morphological differences between the populations ought to be accounted for. To this end, CHIME/FRB is planning a dedicated injection campaign incorporating multi-component bursts and downward-drifting spectral structure. The injection campaigns conducted to date, which include the publicly released sample of nearly $85{,}000$ events presented by \citet{2023AJ....165..152M} and a larger campaign supporting Cat2 population analyses (McGregor et al. in prep.), do not robustly forward-model repeater-like burst morphologies. A population-focused injection framework specifically designed for the observed repeating FRB population is therefore currently under development.

\subsection{Correlations} \label{sec:discussion:correlations}

We do not find any statistically significant correlations among the repeater-level properties considered here; including burst rate, average burst duration, average burst fluence, and extragalactic dispersion measure (DM) which is consistent with previous studies (\citetalias{RN3} \citeyear{RN3}; \citealt{alice}). This absence disfavors simple unified relations between observables (e.g., width–rate or fluence–rate relations) across the repeating FRB population. However, our sample size may still be insufficient to detect weaker trends. Larger, more uniformly characterized samples will therefore be required to robustly test for such subtle or higher-dimensional correlations \citep{connor_2019MNRAS.487.5753C}.

\subsection{DM Variation}
\label{sec:deltadmdisc}
Repeating FRB sources allow us a unique look into the evolution of the environment of the FRB sources as well as their emission over years-long timescales through the multiple measurements of their burst observables. In Figure \ref{fig:deltadm} we present four detections of possible monotonic, linear DM variations in repeating FRBs spanning $-2.05\pm0.11$ to $1.02 \pm 0.05\, \dmunits{}$ year$^{-1}$. Although the presence of linear variability can be considered as established as is suggested by the $p$-values for each source, the rate of change in DM and associated uncertainty of these variations should be treated as rough estimates owing to the degeneracy between repeater-like burst morphology and the DMs we report given the lower time resolution offered by our total intensity data.

Repeaters, on average, are more narrow-band and longer in duration than apparent one-offs. It is well established that repeater burst emission often drifts in frequency; and in particular, much more commonly drifting such that the signal marches down in frequency than upwards \citep{2019ApJ...876L..23H,pleunismorph}. Underlying broader burst components with residual frequency drifting have been observed even when bursts are dedispersed with simultaneous resolved, short-duration (tens of microseconds), broadband emission, for example as shown with exceptionally bright bursts from one of the most prolific repeaters, FRB~20220912A by \cite{2023MNRAS.526.2039H}. 

This kind of morphology, when observed in the regime of low S/N and low time resolution, manifests as diffuse, blobby structures extending to lower frequencies in time in a way that is degenerate with the dispersion due to ionized media along the line of sight of the source, that manifests as a delay proportional to the inverse square of frequency. By over-dedispersing these downward drifting substructures, one can artificially increase the S/N, and thus in this regime, particularly when using S/N maximizing algorithms, the absolute scaling of the DMs is biased high. Without sharp features in the dynamic spectrum that can be reliably used to structure maximize DMs, it is difficult to measure DM values either precisely or accurately (see, for example, the clear upwards bias in DMs for FRB~20220912A in Figure \ref{fig:deltadm}). 

With a large sample of bursts observed at high-time resolution, one can select the high S/N bursts with discernible sharp features such that a structure optimizing DM algorithm will mitigate the effect of this upwards bias to some degree, and would be the most reliable to trace the DM over time. However, despite this upwards bias in the S/N maximized DMs we report here, we do not expect a bias in the evolution of the DM as long as the expected distribution of frequency drifting extents does not evolve with time. If this assumption is false, and the drifting does indeed vary secularly with time, it would have implications for the origin of the drifting, which is not yet understood. We will confirm this either way using our raw voltage data in an upcoming work (Mulyk et al. in prep), as well as any additional bursts we have detected beyond the Cat2 period.

These sources contribute significantly to the total number of known deterministically varying DM FRBs as there are only currently a handful known. We note that two of the candidate's DMs are increasing, and the other two decreasing with time. Such increases have only been seen in thus far in three other repeaters, and usually are observed to be shallower and/or span shorter durations \citep[the DM of FRB~20121102A increased at an average rate of $\sim 0.03\,\dmunits{}$ year$^{-1}$ for six years, FRB~20240619D increased at a rate of $2.4\,\dmunits{}$ year$^{-1}$  for four months, and FRB~20220912A $1.4\pm 0.6\,\dmunits{}$ year$^{-1}$ for about one year; ][]{2019ApJ...876L..23H,2025arXiv251011352S,NIU202676, 2024arXiv241017024O, thomasr117}. 

Upper limits have been placed on the DM variation for a number of other sources.
\cite{2023MNRAS.520.2281N} constrain DM variation from FRB~20200120E to less than 0.15 $\dmunits{}$ year$^{-1}$ over their 10 months of observation. 
This is perhaps not surprising given FRB~20200120E is located in a globular cluster \citep{2021ApJ...910L..18B,2022Natur.602..585K}, and we do not expect globular clusters to contribute a significant amount of internal dispersion \citep{2001ApJ...557L.105F,2007ASPC..365..265R}.
For FRB 20180916B, \cite{2023ApJ...956...23S} constrain DM variation to less than $\sim 1~\dmunits{}$ in their $3+$ years of observation, despite an observed secular RM increase of $\sim 50$ rad m$^{-2}$ (representing a more than 40\% fractional change) during nine months of that same time span \citep{2023ApJ...950...12M}. \cite{2025ApJ...982..154N} present evidence for RM variation in the range of $50-100$ rad m$^{-2}$ for FRB~20190303A and FRB~20200929C. \cite{alice} constrain the DM variation of those same sources to less than $\sim1~\dmunits{}$ over a time period of about five years and one year, respectively. 
While it is possible that in any of the three cases that the lack of DM variation suggests an evolution of the parallel magnetic field rather than the electron density along the line of sight (LOS), the latter cannot be ruled out unambiguously given our conservative upper limit on the DM variation. 
Assuming that both DM and RM variations arise from electron column variability in common magnetoionic plasma and not a changing LOS magnetic field strength, an RM variation of $\sim100~\mathrm{rad}~\mathrm{m}^{-2}$ with a corresponding DM variation of $\lesssim 1~\mathrm{pc}~\mathrm{cm}^{-3}$ necessitates that the average LOS parallel magnetic field strength, $\left|\left<B_\parallel\right>\right|$, be greater than about $120~\mu\mathrm{G}$ using the equation 
\begin{equation}
    \left|\left<B_\parallel\right>\right| = 1.232~\mu\mathrm{G} \left( \frac{\text{RM}}{\mathrm{rad}~\mathrm{m}^{-2}}\right) \left( \frac{\text{DM}}{\mathrm{pc}~\mathrm{cm}^{-3}}\right)^{-1}
\end{equation}
 \cite[see, e.g., ][]{pandhi26}. Note that we have not applied a redshift correction here to convert to the host frame, as the cosmological redshifts of these FRB sources are not all currently known. These magnetic fields strengths (above $\sim 10^2~\mu\mathrm{G}$) are larger than the typical field strength in the ambient interstellar medium of Milky Way-like galaxies \citep{2003astro.ph.10287B,2013pss5.book..641B}. The magnetic fields in pulsar wind nebulae (PWNe) and supernova remnants are expected to have magnetic fields of order $\sim 10-1000 \mu$G, depending on their ages and environments \citep{2012SSRv..166..231R}, and there have been direct measurements of supernova remnant magnetic fields in dense/complicated regions that tend to be even higher than this (see e.g., \citealt{2000ApJ...537..875B}).

In, e.g., the expanding supernova remnant model, which was invoked by \cite{2019ApJ...876L..23H,NIU202676} and \cite{pandhi26} to explain observed DM and RM variation of repeaters, \cite{piro_gaensler} calculate that either positive or negative DM variation on this scale is possible from an expanding supernova remnant. A DM decrease can be expected from a supernova ejecta expanding into an interstellar medium, and it is possible to observe an increase in DM later in the evolution of the system as the supernova remnant sweeps up material. Polarimetric studies of our four varying sources should be particularly constraining for this scenario, given \cite{piro_gaensler} find the local RM should always decrease in proportion to the DM changes if they originate in the same media. Notably, \cite{2023ApJ...951...82M} report an increase of $\sim$ 9 rad m$^{-2}$ between the RMs of the first and third detected bursts of one of our new linear DM variability candidates, FRB~20190117A (the only two bursts from this source with a published RM). Further RM measurements would help to diagnose if this is a deterministic trend.

Regardless, we expect the variation for these four repeaters to be local to the source, as interstellar (Galactic or host), circumgalactic, and intergalactic media are not expected to evolve on this scale over years-long timescales \citep{2013MNRAS.435.1610P,2017ApJ...841..125J,pandhi26}. For FRB 20121102A, \cite{2017ApJ...847...22Y} show that expected DM variations due to large-scale effects are negligibly small, and only the local effects they explore, such as those due to an expanding supernova remnant, a PWN, motion within or growing HII regions, or plasma lenses could create this scale of variation.

There are no systems which exhibit truly analogous behavior to this in our Galaxy, that is, demonstrate the magnitude of DM variation on the timescale suggested by these observations. Pulsars in binary systems with large main sequence companions (e.g., a Be star) would have sufficiently dense surrounding plasma required to invoke the observed magnitude of DM changes, like PSR B1259--63 \citep{10.1093/mnras/279.3.1026} or PSR~J2108+4516 \citep{2023ApJ...943...57A}, but the phase-locked DM variability of these sources occurs in a small fraction of their orbits, so a much longer orbit observed at a finely tuned phase would be required to approximate linearity across the years-long FRB observations. 
It is possible that the Galactic center magnetar, PSR J1745$-$2900, is exhibiting linear DM variations on approximately these scales \citep{2018ApJ...852L..12D}. This DM evolution would be accompanying the source's significant and, at times,  approximately linear RM variation, but the DM measurements of PSR J1745--2900 have non-neglible uncertainties and hence the DMs are still consistent with being constant in time at the 2$\sigma$ level \citep{2018ApJ...852L..12D}.  

DM variations, both increasing and decreasing, have been observed for the Crab pulsar that are smaller and shorter duration ($\lesssim 10^{-1}\, \dmunits{}$ over a few weeks).  It is thought these variations are due to plane of sky movement of filamentary structure of the PWN along our LOS, rather than due to the evolution of the PWN itself \citep{2008A&A...483...13K,2018MNRAS.479.4216M,2025arXiv250723201S}. 
Given the smaller variation and shorter duration of its monotonicity, a younger and/or more extreme pulsar/PWN system would likely need to be invoked to explain the observed variation. It is worth noting that radio luminosities 10 times brighter than that of the Crab PWN have been ruled out for FRB~20250316A (\citetalias{rbfloat} \citeyear{rbfloat}) and there has not yet been any direct association between an observed supernova and a FRB source \citep{2025ApJ...991..199D}.

\section{Conclusion}
We report uniform population statistics of 80 repeating FRBs, including the discovery of \nrepeaters{} from CHIME/FRB's Second catalog. This increases the total published repeater count to 91 (\citealt{R1}; \citetalias{R2} \citeyear{R2};   \citetalias{RN1} \citeyear{RN1}; \citealt{2019ApJ...887L..30K};\citealt{RN2, 2021ApJ...910L..18B, 2020Natur.586..693L,2022ATel15679....1M,2022Natur.606..873N,2022ApJ...927...59L};  \citetalias{RN3} \citeyear{RN3};\citealt{2023ATel16191....1R,2024ATel16734....1A,2025MNRAS.540.1685T,2025ApJ...979L..21S, 2025arXiv250513297S,2025ATel17548....1T}) 
and represents the largest sample of repeating FRBs yet, detected by a well-characterized and homogeneous survey, ideal to yield meaningful inference from full population syntheses. 

We estimate burst rates above 5 Jy ms for these new repeaters that span $10^{-5.7}$ to $10^{-0.5}$ bursts hr$^{-1}$. The extragalactic DMs, assuming the NE2001 model \citep{ne2001}, range between $99.4$ and $1446.0\ \dmunits$. There are a number of notable sources included in this sample, like FRB~20210601A, which seems to be an extreme outlier in its observed peak rate compared to the duration of its known activity (Mckinven et al. in prep.). Also included in this sample is FRB~20220529A, the DM of which seems to be continually decreasing over our years of observation, with the notable exception of a brief increase just before its extreme RM flare \citep{2025arXiv250304727L,pandhi26}. We present an additional four repeaters that exhibit DM variation, and this DM variation appears to be monotonic and linear on years-long timescales, with rates of change spanning --2.05 to 1.02 $\dmunits{}$ year$^{-1}$, although we caution these measurements are based on our lower-resolution total intensity data, which is known to be biased towards higher DMs. We will revisit the variation for these sources using our raw voltage data and detections beyond the Cat2 timeframe, where available, in a future work, in addition to analyzing their polarimetric and scattering properties. 

While our observed repeater burst rates are, on average, slightly higher than the upper limits on burst rates implied by our observations of as-yet one-off FRBs, the upper limits are entirely contained within the range of observed repeater rates, and the distribution of repeat rates from all FRBs does not appear bimodal. Thus, from rates alone we find no evidence for two distinct populations.

The fraction of FRBs that have been observed to repeat does not evolve significantly with observation duration. We measure the observed repeater fraction to be $2.4 \pm 0.4\%$, which is a lower limit on the true fraction of FRBs that repeat. In the continuum scenario of FRB rates, this number is only informative of the fraction of FRBs which are bright enough to be detected twice with our survey.

While we have yet to find evidence of bimodality in the FRB population beyond the observed phenomenological differences \citep{pleunismorph,alice}, the increasing sample sizes of repeaters and as-yet one-offs from uniform surveys like CHIME/FRB provide a promising avenue for discovery. This could become particularly meaningful with foreseen corrections for the bias against repeater-like morphology from future injection analyses, a bias which is not accounted for in this work.

Other promising venues for determining whether there exist more than one underlying population of FRBs are upcoming via large samples of precise localizations being available not only for repeating FRBs but also one-off FRBs from ASKAP, CHIME/FRB Outriggers, DSA-110, and MeerKAT currently. These localizations will allow for large samples of FRBs with known distances, hence decreasing uncertainty in population models.  The similarity of properties of the host galaxies and local environments of repeating and one-off FRB samples could be additionally illuminating for whether the populations are likely to share a common progenitor. Upcoming surveys that will increase sample numbers by orders of magnitude are CHORD, DSA, and the SKA.

\begin{acknowledgments}

We are grateful to Clancy James for numerous useful discussions and ideas, and also assistance with using the \texttt{zDM} package, all of which improved the quality of the work. 
We acknowledge that CHIME is located on the traditional, ancestral, and unceded territory of the Syilx/Okanagan people. We are grateful to the staff of the Dominion Radio Astrophysical Observatory, which is operated by the National Research Council of Canada. CHIME operations are funded by a grant from the NSERC Alliance Program and by support from McGill University, University of British Columbia, and University of Toronto. CHIME was funded by a grant from the Canada Foundation for Innovation (CFI) 2012 Leading Edge Fund (Project 31170) and by contributions from the provinces of British Columbia, Québec and Ontario. The CHIME/FRB Project was funded by a grant from the CFI 2015 Innovation Fund (Project 33213) and by contributions from the provinces of British Columbia and Québec, and by the Dunlap Institute for Astronomy and Astrophysics at the University of Toronto. Additional support was provided by the Canadian Institute for Advanced Research (CIFAR), the Trottier Space Institute at McGill University, and the University of British Columbia. The CHIME/FRB baseband recording system is funded in part by a CFI John R. Evans Leaders Fund award to IHS.

A.M.C. is a Banting Postdoctoral Research Fellow.
D. B. holds an NSERC Discovery Grant and a CANSSI Collaborative Research Team Grant that helped support this work.
A.P.C. is a Canadian SKA Scientist and is funded by the Government of Canada / est financé par le gouvernement du Canada.
G.M.E. holds an NSERC Discovery Grant and a CANSSI Collaborative Research Team Grant that helped support this work.  J.W.T.H., Z. P., and the AstroFlash research group at McGill University, University of Amsterdam, ASTRON, and JIVE are supported by: a Canada Excellence Research Chair in Transient Astrophysics (CERC-2022-00009); an Advanced Grant from the European Research Council (ERC) under the European Union’s Horizon 2020 research and innovation programme (`EuroFlash’; Grant agreement No. 101098079); an NWO-Vici grant (`AstroFlash’; VI.C.192.045); an NSERC Discovery Grant (RGPIN-2025-06681); an ERC Starting Grant (`EnviroFlash’; Grant agreement No. 101223057); and an NWO-Veni grant (VI.Veni.222.295).
C. L. acknowledges support from the Miller Institute for Basic Research at UC Berkeley.
A.P. is a Trottier Space Institute Postdoctoral Fellow.
P.S. acknowledges the support of an NSERC Discovery Grant (RGPIN-2024-06266).
D.C.S. holds an NSERC Discovery Grant (RGPIN-2021-03985) and received prior support from a CANSSI Collaborative Research Team Grant.
F.A.D. is a Canadian SKA scientist and is funded by the National Research Council of Canada. / est financé par le Conseil national de recherches du Canada.
E.F and S.S.P. are supported by the National Science Foundation under grant AST-2407399.
M.L. acknowledges the support of the Natural Sciences and Engineering Research Council of Canada (NSERC-CGSD). 
V.M.K. holds the Lorne Trottier Chair in Astrophysics \& Cosmology, a Distinguished James McGill Professorship, and receives support from an NSERC Discovery grant (RGPIN 228738-13) and is grateful for hospitality as part of a Visiting Professorship at Tel Aviv University.
K.W.M. is supported by NSF Grant No. 2510771 and receives lumbar support from the Adam J. Burgasser Chair in Astrophysics.
K.T.M. is supported by a Fonds de Recherche du Quebec—Nature et Technologies (FRQNT) Master’s Research Scholarship. 
D.M. acknowledges support from the French government under the France 2030 investment plan, as part of the Initiative d'Excellence d'Aix-Marseille Universit\'e -- A*MIDEX (AMX-23-CEI-088).
M.N. is a FRQNT postdoctoral fellow. 
K.N. acknowledges support by NASA through the NASA Hubble Fellowship grant \# HST-HF2-51582.001-A awarded by the Space Telescope Science Institute, which is operated by the Association of Universities for Research in Astronomy, Incorporated, under NASA contract NAS5-26555.
A.B.P. acknowledges support by NASA through the NASA Hubble Fellowship grant HST-HF2-51584.001-A awarded by the Space Telescope Science Institute, which is operated by the Association of Universities for Research in Astronomy, Inc., under NASA contract NAS5-26555. A.B.P. also acknowledges prior support from a Banting Fellowship, a McGill Space Institute~(MSI) Fellowship, and a FRQNT Postdoctoral Fellowship. 
M.W.S is a FRQNT postdoctoral fellow and acknowledges support from the Trottier Space Institute Fellowship program.
V.S. is supported by a FRQNT Doctoral Research Award.
FRB research at UBC is supported by an NSERC Discovery Grant and by the Canadian Institute for Advanced Research.

\end{acknowledgments}

%

\vspace{5mm}


\appendix

\begin{deluxetable*}{lllllllllll}
\renewcommand\thetable{A}
\tablehead{
\colhead{Source} & \colhead{$P_{\text{CC}}$} & \colhead{R.A.\tablenotemark{a}} & \colhead{Decl.\tablenotemark{a}}  & \colhead{DM} & \colhead{$\text{DM}_{\text{NE2001}}$} & \colhead{$\text{DM}_{\text{YMW16}}$} & \colhead{$N$\tablenotemark{b}} & \colhead{Exposure} & \colhead{Sensitivity\tablenotemark{c}} & \colhead{Rate $ \geq \text{5 Jy ms}$\tablenotemark{d}}\\ 
\colhead{} & \colhead{} & \colhead{(deg)} & \colhead{(deg)}  & \colhead{(pc cm$^{-3}$)} & \colhead{(pc cm$^{-3}$)} & \colhead{(pc cm$^{-3}$)} & \colhead{} & \colhead{(U/L, hr)} & \colhead{(U/L, Jy ms)} & \colhead{(hr$^{-1}$)}}
\tablecaption{Properties of \nsilver{} New Candidate Repeaters (Our Silver Sample)\label{tab:silverprop}}
\startdata
FRB~20200905B & $4\times 10^{-5}$ & 336.60(8) & 72.7(2) & 305.53(5) & 162.6 & 198.0 & 2 & 300(100)/260(40) & 3.4/24.0 & $3_{-3}^{+9}\times 10^{-3}$ \\
FRB~20210219C & $5\times 10^{-5}$ & 350.417(4) & 87.984(4) & 526.262(3) & 469.0 & 467.8 & 2 & 2100(100)/2950(30) & 582.9/39.4 & $0.4_{-0.2}^{+1.0}$ \\
FRB~20210305A & $6\times 10^{-5}$ & 208.61(4) & 76.8(1) & 254.417(8) & 221.4 & 214.2 & 2 & 420(40)/489.9(1) & 0.04 & $1.3_{-0.8}^{+5.0}\times 10^{-6}$ \\
FRB~20181018B & $7\times 10^{-5}$ & 336.72(5) & 71.71(1) & 293.813(3) & 137.0 & 179.2 & 2 & 50(50)/257(2) & 11.3 & $0.5_{-0.4}^{+5.0}\times 10^{-1}$ \\
FRB~20201226B & $1\times 10^{-4}$ & 130.306(6) & 4.828(6) & 282.43(10) & 225.2 & 223.8 & 2 & 86(3) & 11.3 & $3_{-1}^{+10}\times 10^{-2}$ \\
FRB~20200510A & $1\times 10^{-4}$ & 143.894(4) & 36.409(4) & 290.878(3) & 264.0 & 251.0 & 2 & 111(4) & 6.9 & $2_{-2}^{+5}\times 10^{-2}$ \\
FRB~20191102B & $1\times 10^{-4}$ & 182.155(6) & 86.902(6) & 598.35(1) & 553.0 & 548.1 & 2 & 2007(7)/400(100) & 1.0/62.7 & $3_{-1}^{+10}\times 10^{-5}$ \\
FRB~20211006B & $2\times 10^{-4}$ & 134.064(5) & 73.227(6) & 310.219(1) & 270.5 & 263.8 & 2 & 300(20)/207(9) & 1.3/9.3 & $0.7_{-0.7}^{+1.0}\times 10^{-3}$ \\
FRB~20200125B & $2\times 10^{-4}$ & 251.437(5) & 53.094(5) & 512.400(2) & 480.9 & 473.4 & 2 & 90(20) & 1.5 & $1_{-1}^{+6}\times 10^{-3}$ \\
FRB~20181122B & $2\times 10^{-4}$ & 308.464(8) & 84.635(8) & 225.7112(8) & 167.8 & 166.5 & 2 & 400(100)/1050(30) & 0.07/0.1 & $0.3_{-0.2}^{+1.0}\times 10^{-5}$ \\
FRB~20230629A & $2\times 10^{-4}$ & 339.332(5) & 30.781(5) & 303.835(3) & 253.7 & 245.3 & 2 & 93(4) & 2.2 & $5_{-5}^{+10}\times 10^{-3}$ \\
FRB~20210313A & $2\times 10^{-4}$ & 203.054(7) & 87.572(7) & 398.794(4) & 352.3 & 347.7 & 2 & 1210(20)/2450(20) & 0.7/1.1 & $0.3_{-0.2}^{+1.0}\times 10^{-4}$ \\
FRB~20191030C & $2\times 10^{-4}$ & 178.679(4) & 61.713(4) & 339.562(4) & 314.9 & 306.0 & 2 & 10(10) & 1.6 & $0.9_{-0.8}^{+7.0}\times 10^{-2}$ \\
FRB~20211102E & $2\times 10^{-4}$ & 14.208(6) & 84.652(6) & 256.436(1) & 185.4 & 189.3 & 2 & 970(40)/1190(20) & 0.02/0.03 & $1.4_{-0.8}^{+5.0}\times 10^{-7}$ \\
\enddata
\tablenotetext{a}{J2000, ICRS}
\tablenotetext{b}{The total number of bursts detected CHIME/FRB by the source. Includes the bursts detected outside of our FWHM at 600 MHz.}
\tablenotetext{c}{Fluence threshold for the source at the 95\% confidence level.}
\tablenotetext{d}{Burst rate corresponding to the upper transit, and scaled to a fluence threshold of 5 Jy ms from the sensitivity in the previous column assuming a powerlaw index of $\alpha = -1.5$.}
\end{deluxetable*}

\begin{deluxetable}{cclll}
\renewcommand\thetable{B}
\tablehead{
\colhead{Source} & \colhead{$N_{\text{bursts}}$\tablenotemark{a}} & \colhead{Exposure\tablenotemark{b}} & \colhead{Sensitivity\tablenotemark{c}} & \colhead{Rate ($\geq 5$ Jy ms)\tablenotemark{d}}\\ 
\colhead{} & \colhead{} & \colhead{(hours)} & \colhead{(Jy ms)} & \colhead{(hr$^{-1}$)}}
\tablecaption{Updated rates in the upper transit for select previously published CHIME/FRB repeaters \citep{2022ApJ...927...59L,2023ApJ...956...23S,2022ATel15679....1M}. \label{tab:extrarates}}
\startdata
FRB~20180916B & 55 &295.2 & 5.2 & $2.0^{+0.5}_{-0.6}\times10^{-1}$\\
FRB~20201124A & 18 & 92.2 & 12.9&  $8^{+3}_{-4}\times 10^{-1}$\\
FRB~20220912A & 210 & 133.3 & 0.43 & $4.0^{+0.5}_{-0.5}\times 10^{-2}$\\
\enddata
\tablenotetext{a}{Number of bursts detected by CHIME/FRB from this source that fall within the FWHM at 600 MHz of the corresponding detection beam, where our exposure is defined. }
\tablenotetext{b}{Since these sources all have milliarcsecond scale localizations \citep{2020Natur.577..190M,2022ApJ...927L...3N,2024MNRAS.529.1814H}, there is negligible error on their exposure estimates. }
\tablenotetext{c}{Fluence threshold for the source at the 95\% confidence level.}
\tablenotetext{d}{Burst rate corresponding to the upper transit, and scaled to a fluence threshold of 5 Jy ms from the sensitivity in the previous column assuming a powerlaw index of $\alpha = -1.5$.}
\end{deluxetable}

\bibliography{bib}{}
\bibliographystyle{aasjournal}



\end{document}